\documentclass{aastex61}
\usepackage{amsmath}
\usepackage{natbib}
\begin{document}

\title{An Analytic Formulation  of 21-cm Signal from Early Phase of Epoch of Reionization}

\author{Janakee Raste}
\thanks{Joint Astronomy Program, Indian Institute of Science, Bangalore 560012, India}
\affiliation{Raman Research Institute, 
Bangalore 560080, India} 

\author{Shiv Sethi}
\affiliation{Raman Research Institute, 
Bangalore 560080, India}

\begin{abstract}
We present an analytic formulation to model the fluctuating component 
of the \ion{H}{1} signal from the epoch of 
reionization during the phase of partial heating. During this phase, 
we assume  self-ionized regions,  whose size distribution can be computed 
using excursion set formalism, to be surrounded by heated regions. We model
the evolution of heating profile around these regions (near zone) and  their merger into the time-dependent  background (far zone). We develop a formalism 
to compute the two-point correlation function for this topology, taking into
account the heating auto-correlation and heating-ionization cross-correlation.
We model the ionization and X-ray heating   using four parameters: efficiency of 
ionization, $\zeta$, number of X-ray photons per stellar baryon, $N_{\rm heat}$, 
the spectral index of X-ray photons, $\alpha$, and the minimum frequency of 
X-ray photons, $\nu_{\rm min}$. We compute the \ion{H}{1} signal in the redshift range
$10 < z < 20$ for the $\Lambda$CDM model for a set  of these parameters. 
We show that the \ion{H}{1} signal  for a range of scales $1\hbox{--}8 \, \rm Mpc$  show a peak strength 
$100\hbox{--}1000 \, \rm (mK)^2$ during the partially heated era. The 
redshift at which the signal makes a transition to uniformly heated 
universe  depends on modelling parameters, e.g. if $\nu_{\rm min}$ is changed 
from $100 \, \rm eV$ to $1 \, \rm keV$, this transition moves from $z \simeq 15$ to $z \simeq 12$.  This 
result,  along with the dependence of the \ion{H}{1} signal on modelling parameters, 
is in reasonable agreement with existing results from  N-body simulations.

\end{abstract}


\section{Introduction} \label{sec:intro}
The probe of the epoch of reionization remains one of the outstanding aims
of modern cosmology. In the past 15 years, important strides have been made in this 
direction, mainly led by the detection of Gunn-Peterson effect at $z \simeq 6$
and the CMB temperature and polarization anisotropy detections by WMAP and Planck (\cite{Ade:2013zuv,Hinshaw:2012aka,Ade:2015xua,Fan:2000gq,2001AJ....122.2850B}). The former discovery
suggests that the universe could be making a transition from neutral to fully 
ionized at $z\simeq 6$,  while the latter shows the universe might have been
fully ionized at $z \simeq 8.5$. The current best bounds on the reionization optical depth from Planck put stringent constraints on  the redshift of reionization:  $z_{\rm reion} = 8.5 \pm 1$ (\cite{Ade:2015xua}).

One important missing piece in these probes is that neither seem capable of 
discerning the dynamics of the reionization process. For instance, the CMB 
anisotropies (both polarization and temperature) are sensitive only to the integrated optical depth through the surface of reionization (e.g. \cite{2003moco.book.....D}). 

Theoretical estimates
suggest that dark age of the universe might have ended around $z \simeq 30$
with the formation of first structures in the universe. These 
structures are expected to emit  UV light which might have reionized the 
universe  at $z \simeq 9$ (see e.g. \cite{Barkana:2000fd,21cm_21cen,2014PTEP.2014fB112N,2010ARA&A..48..127M} and references therein). The most direct way to probe this transition is through  the detection of redshifted hyperfine 21~cm line 
of neutral hydrogen (\ion{H}{1}). 
The past one decade has seen  major progress on both theoretical
and experimental front  in this endeavour. Currently, there are many ongoing 
experiments that are attempting to detect both the global \ion{H}{1} signal and its 
fluctuating component from the epoch of reionization. The \ion{H}{1} signal 
from the epoch has been computed  both analytically  and  using numerical
simulations. Theoretical estimates show that the global signal
is observable in both absorption and emission with its strength in the 
range $-200\hbox{--}20 \, \rm mK$ in a frequency range of $50\hbox{--}150 \, \rm MHz$, which corresponds roughly to a redshift range $25 > z > 8$ (e.g. \cite{1997ApJ...475..429M,2000NuPhS..80C0509T,2004ApJ...608..611G,Sethi05}).  The fluctuating component of the signal is likely to be an order of magnitude smaller 
on scales in the range $3\hbox{--}100 \, \rm Mpc$, which implies  angular scales
in the range $\simeq 1\hbox{--}30$~arc-minutes (e.g. \cite{ZFH04,FZH04a,FZH04b,2007MNRAS.376.1680P}; for comprehensive reviews see e.g. \cite{21cm_21cen,2014PTEP.2014fB112N,2010ARA&A..48..127M}). Many of the ongoing and upcoming experiments have the the capability to detect this signal in hundreds of hours of integration (e.g. \cite{2015aska.confE...3A,2014MNRAS.439.3262M}). Upper limits on the fluctuating 
component of the \ion{H}{1}  signal have  been obtained by many ongoing experiments---GMRT, MWA, PAPER, and LOFAR (\cite{2017ApJ...838...65P,2016ApJ...833..102B,2015ApJ...809...61A,2013MNRAS.433..639P}). The best current upper limits corresponds to: $k^3 P(k)/(2\pi^3) < (22.4 \,\rm mK)^2$ for $0.15 < k < 0.5 \, \rm h Mpc^{-1}$ at $z \simeq 8.4$ (\cite{2015ApJ...809...61A}).

The \ion{H}{1}  signal carries crucial information about the first sources in the universe. In particular, the \ion{H}{1} signal is determined by the radiation emitted by these 
sources in three frequency bands: UV radiation that ionizes the medium, 
Lyman-$\alpha$ radiation (frequencies between Lyman-limit and Lyman-$\alpha$), 
and X-ray photons (all photons with energies much higher than hydrogen and helium ionization threshold). The emission in these three bands determines  the 
evolution of the global \ion{H}{1} signal. And, along with primordial density perturbations given by the $\Lambda$CDM model,  the perturbations  of these radiation
fields establish the length  scales of  the fluctuating component of the signal. 

In this paper, our main focus is the analytic modelling of the fluctuating 
component of the \ion{H}{1} signal in its early phase when the universe is partially
heated.  This phase of the EoR has 
been extensively studied using numerical methods and analytic estimates (e.g. \cite{2007MNRAS.376.1680P,2013MNRAS.435.3001T,2013MNRAS.431..621M,LateHeat2,2015MNRAS.447.1806G,2014MNRAS.443..678P,2017MNRAS.464.3498F,2012Natur.487...70V,21CMFAST}). 
We present a new formalism in this paper which seeks to unravel the 
correlation structure of the fluctuations  based on the topology of 
the ionization and heating regions. 

In the next section, we review the \ion{H}{1} signal from the epoch of reionization
and discuss  the impact of three radiation fields on the signal. In particular,
the modelling of X-ray heating is described in detail. In section~\ref{autoh1sig}, we 
present our formalism for  computing the two-point correlation function of the \ion{H}{1} 
signal. We also discuss various approximations, assumptions, and limits germane 
to our formulation. In section~\ref{sec:res}, we present  our results and compare them with the  inferences  of other studies. In section~\ref{sec:sumcon} we summarize our findings and make concluding remarks. Throughout this paper, 
we assume the spatially-flat $\Lambda$CDM model with the following 
parameters: $\Omega_m = 0.254$, $\Omega_B = 0.049$, $h = 0.67$ and $n_s = 0.96$, 
with the overall normalization corresponding to  $\sigma_8 = 0.83$ (\cite{Ade:2015xua}). 

\pagebreak

\section{Cosmic Dawn and Epoch of Reionization}\label{sec:h1sig}
In the  rest frame, hyperfine splitting of the ground state of neutral hydrogen
causes an energy difference that corresponds to a wavelength $\lambda = 21.1 \, \rm cm$. The excitation temperature of this line, $T_S$, is determined  by three processes in early universe: emission and 
absorption of  CMB photons which is a blackbody of temperature $T_{\rm CMB}$, 
collisions with atoms, and the mixing of the two levels caused 
by  Lyman-$\alpha$ photons (Wouthuysen-Field effect).  $T_S$ can be expressed in terms of 
the colour temperature of Lyman-$\alpha$ photons, $T_{\alpha}$,   gas kinetic temperature $T_K$, and $T_{\rm CMB}$ (\cite{Field1958,21cm_21cen}):
	\begin{equation}
		T_S=\frac{T_{\rm CMB}+y_{\alpha}T_{\alpha}+y_c T_K}{1+y_{\alpha}+y_c}
                \label{eq:tsbas}
	\end{equation}
Here $y_c \propto n_{\rm H}$ and $y_\alpha \propto n_\alpha$ ($n_{\rm H}$ and $n_\alpha$
are the number densities of neutral hydrogen atoms and Lyman-$\alpha$ photons,
respectively) determine the efficiency of  collisions and Lyman-$\alpha$ photons, respectively. In the early universe, $1000 < z < 100$, $T_S$ relaxes to $T_{\rm CMB}$. In the redshift range $100 < z < 30$, collisions determine the spin temperature and $T_S$ relaxes to the kinetic 
temperature $T_K$ of the matter. As the epoch of reionization commences, the production of Lyman-$\alpha$ photons couples the spin temperature to the colour 
temperature of Lyman-$\alpha$ $T_\alpha$. 
It can be shown that multiple scattering of Lyman-$\alpha$ photons with \ion{H}{1}  causes $T_\alpha$ to relax to the kinetic temperature (e.g. \cite{2004ApJ...602....1C,1959ApJ...129..551F,1994ApJ...427..603R}). 
Therefore if $y_{\text{tot}}=y_c+y_{\alpha} \gtrsim T_{\rm CMB}/T_K $, then $T_S$ is strongly coupled to $T_K$. Otherwise, it relaxes to $T_{\rm CMB}$. 

The \ion{H}{1} emits or absorbs 21-cm radiation from CMB depending on whether its $T_S$ is greater than or less than $T_{\text{CMB}}$. This temperature difference is observable and can be expressed as (e.g. \cite{21cm_21cen,1997ApJ...475..429M,1999A&A...345..380S,2004ApJ...608..611G,Sethi05}):
	\begin{align}
		\Delta T_b  &=\frac{1-\mathrm{e}^{-\tau}}{1+z}(T_S-T_{\text{CMB}}) \nonumber \\
				& \simeq \frac{\tau}{1+z}(T_S-T_{\text{CMB}}) \nonumber \\
				& \simeq 26.25\;n(1+\delta)\left(1-\frac{T_{\text{CMB}}}{T_S}\right) \left(\frac{1+z}{10}\frac{0.14}{\Omega_m h^2}\right)^{\frac{1}{2}} \left(\frac{\Omega_b h^2}{0.022}\right) \text{mK} \label{overallnorm}
	\end{align}
Here we ignore redshift space distortion. In writing the expression, we 
have expressed the \ion{H}{1} number density as, $n_H = \bar n_H n (1+\delta)$; $\delta$ corresponds to density inhomogeneities in the gas. 
The value of  mean density $\bar n_H$ has been absorbed in the prefactor of 
Eq.~(\ref{overallnorm}). In our formulation every small volume  is either completely neutral or completely 
ionized, therefore we define a variable $n$ which is unity if the medium is neutral and zero otherwise.   In addition to density inhomogeneities, there are ionization and 
spin temperature, $T_S$,  inhomogeneities (as the medium is partially ionized or partially heated).  
We further  define dimensionless temperature fluctuation as (\cite{ZFH04}):
	\begin{align}
		\psi &= n(1+\delta)\left(1-\frac{T_{\text{CMB}}}{T_S}\right) \nonumber \\
			&= n(1+\delta)(1-s) \label{psidef}
	\end{align}
We have defined  $s = T_{\text{CMB}}/T_S$. The statistics of  $\psi$ allows
us to study the main physical processes that cause brightness temperature fluctuations: $\delta$ (density perturbations), $n$ (ionization inhomogeneities) and 
$s$ (fluctuations of spin temperature). All these quantities
are functions of   position in space but we suppress this dependence for 
notational clarity. In this paper, we assume $T_\alpha = T_K$ everywhere, as discussed in section~\ref{sec:lyalpha}. 

We next consider the impact of the three  radiation fields---ionizing radiation, Lyman-$\alpha$ photons, and X-ray photons---on the  brightness temperature 
inhomogeneities.  

\subsection{Photo-ionization} \label{sec:photion}
At the end of the dark ages, high density regions of the universe collapse
and form structures of  different masses. In our work we assume
that the smallest mass that  can collapse corresponds to  \ion{H}{1}-cooled halo (e.g. \cite{Barkana:2000fd}):
    \begin{align}
        M_\text{min} = 3.915 \times 10^8 \frac{\Omega_m^{1/2}}{h\;(1+z)^{3/2}} M_\odot
    \end{align}
These collapsed structures  emit UV photons, which are absorbed in the immediate vicinity of the sources and carves out \ion{H}{2} regions around them in the IGM. These structures also emit  Lyman-$\alpha$ and X-rays radiation  which  penetrates further into the IGM.

The size distribution of the ionization bubbles can be computed using excursion set formalism by defining self-ionized regions (\cite{FZH04a}). These regions  have enough sources to ionize all the gas in it. Such regions are not created by a single source but rather a set of highly clustered
sources which is the case in the early universe for $\Lambda$CDM model,  and therefore these regions are larger than the \ion{H}{2} regions of a single source.  Here we briefly describe the formalism (for details see e.g. \cite{FZH04a} and 
references therein). We start by defining  $\zeta$, the ionization efficiency factor:
	\begin{equation}
        \zeta = f_\star f_{\text{esc}}N_{\text{ion}}N_{\text{rec}}^{-1}
\label{eq:defzeta}
    \end{equation}
Here  $f_\star$ is the fraction of collapsed baryons that is converted into stars. $f_{\text{esc}} $ equals the fraction of ionizing photons that escape the source halo and $N_{\text{ion}}$ is  number of hydrogen ionizing (UV) photons created per stellar baryon while  $N_{\text{rec}}$ is the number of recombinations. We assume
$\zeta$ to be constant in this paper,  even though 
it could be time dependent owing  to the evolution of quantities used to define it.  Inside a self-ionization region, $1/\zeta = f_{\text{coll}}=M_\text{coll}/M_\text{tot}$, where $f_{\rm coll}$ is the 
fraction of collapsed mass inside the self-ionized region. 

Using the extended Press-Schechter model, the collapse fraction can be expressed as,
    \begin{align}
        f_{\text{coll}}=\text{erfc}\left[ \frac{\delta_c(z)-\delta_m}{\sqrt{2[\sigma^2_\text{min}-\sigma^2(m)]}} \right] \nonumber
    \end{align}
Here $\sigma^2(m)$ is the variance of density fluctuations for  mass $m$, 
the mass of the self-ionized region at $z = 0$;  $\sigma^2_\text{min}\equiv \sigma^2(m_\text{min})$ and $\delta_c(z) = 1.68/D_+(z)$ is the critical density for collapse at redshift $z$ and $D_+$ is the growing mode of density perturbations. $\delta_x(m,z)$ is the redshift and mass dependent barrier for excursion set formalism. The linear fit to this true barrier at $m \rightarrow \infty$ is,
	\begin{align}
		B(m,z) 	&= \delta_c(z)-\sqrt{2}\;K(\zeta)\sigma_\text{min} + \frac{K(\zeta)\sigma^2(m)}{\sqrt{2}\sigma_\text{min}} \nonumber
	\end{align}
and $K(\zeta)= \text{erf}^{-1}(1-\zeta^{-1})$. To find the self-ionized region, we need to find the first up-crossing of $\delta$ above the curve described by $B(m,z)$. We can write the mass function analytically as (\cite{1998MNRAS.300.1057S}),
	\begin{align}
		m\frac{\mathrm{d}n}{\mathrm{d}m} = \sqrt{\frac{2}{\pi}} \frac{\bar{\rho}}{m} \left|\frac{\mathrm{d\;ln}\;\sigma}{\mathrm{d\;ln}\;m}\right| \frac{B_0}{\sigma(m)}\text{exp}\left[-\frac{B^2(m,z)}{2\sigma^2(m)}\right] 
\label{eq:massfun}
	\end{align}
Where $B_0 \equiv \delta_c(z)- \sqrt{2}\; K(\zeta) \sigma_\text{min}$ is the value of barrier at $m \rightarrow \infty$.  Eq.~(\ref{eq:massfun}) gives  the comoving number density of self-ionized  bubbles in the mass range $(m,m+\mathrm{d}m)$. Here $\bar\rho$ is the background mass density. We use $\Lambda$CDM power spectrum (the matter power spectrum of the 
$\Lambda$CDM model is generated using publicly-available code CMBFAST) for solving Eq.~(\ref{eq:massfun}). We note that this formalism has been used extensively for analytic 
work and for simulations of epoch of reionization, including in the publicly-available code 21cmFAST( \cite{21CMFAST}). In a numerical simulation, the self-ionized region
is constructed by identifying  the largest contiguous  region  that satisfies the condition
for a self-ionized bubble: $f_{\rm coll} = 1/\zeta$. This region need not be 
spherical. For analytic work, we assume the region to be spherical. We discuss 
the implications of this assumption in a later section. 

\begin{figure} 
	\centering
	\begin{minipage}{0.49\textwidth}
	\centering
		\includegraphics[width=1.0\linewidth]{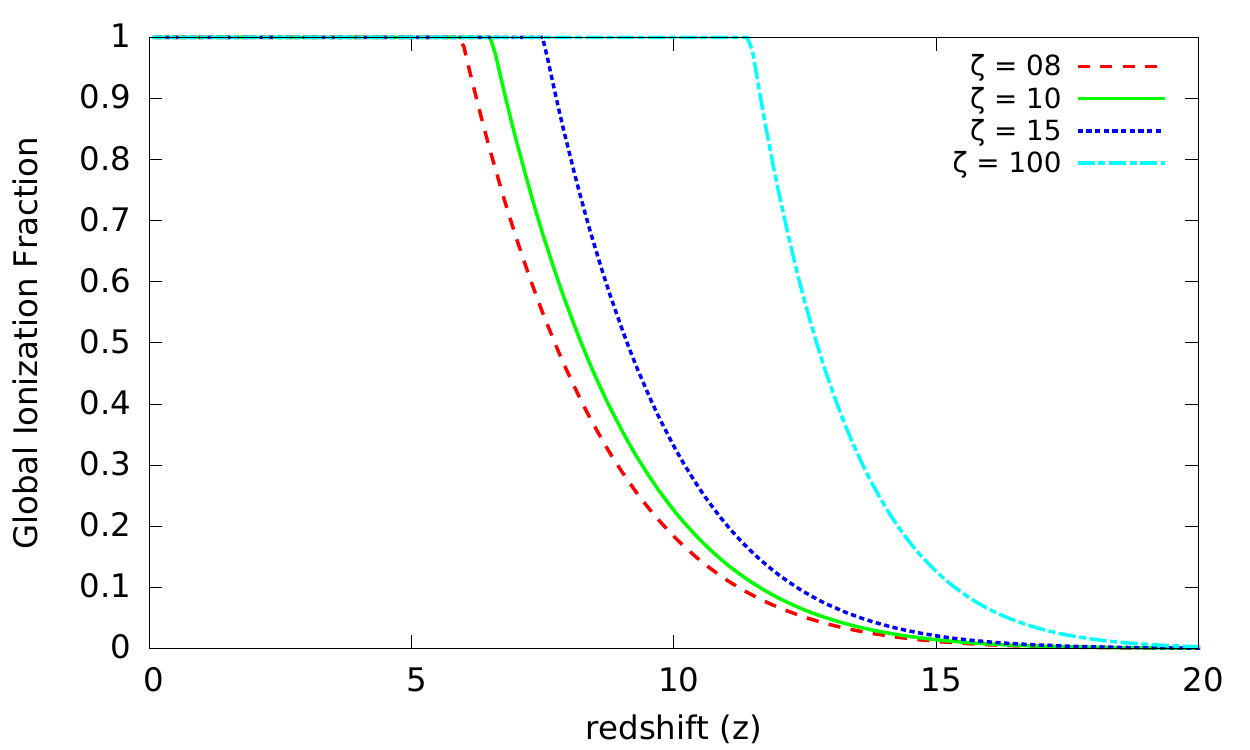}
		\caption{Evolution of global ionization fraction ($f_i$) is 
shown  for different values of $\zeta$.}
		\label{fig:ionzeta}
	\end{minipage}\hfill
	\begin{minipage}{0.49\textwidth}
		\centering
		\includegraphics[width=1.0\textwidth]{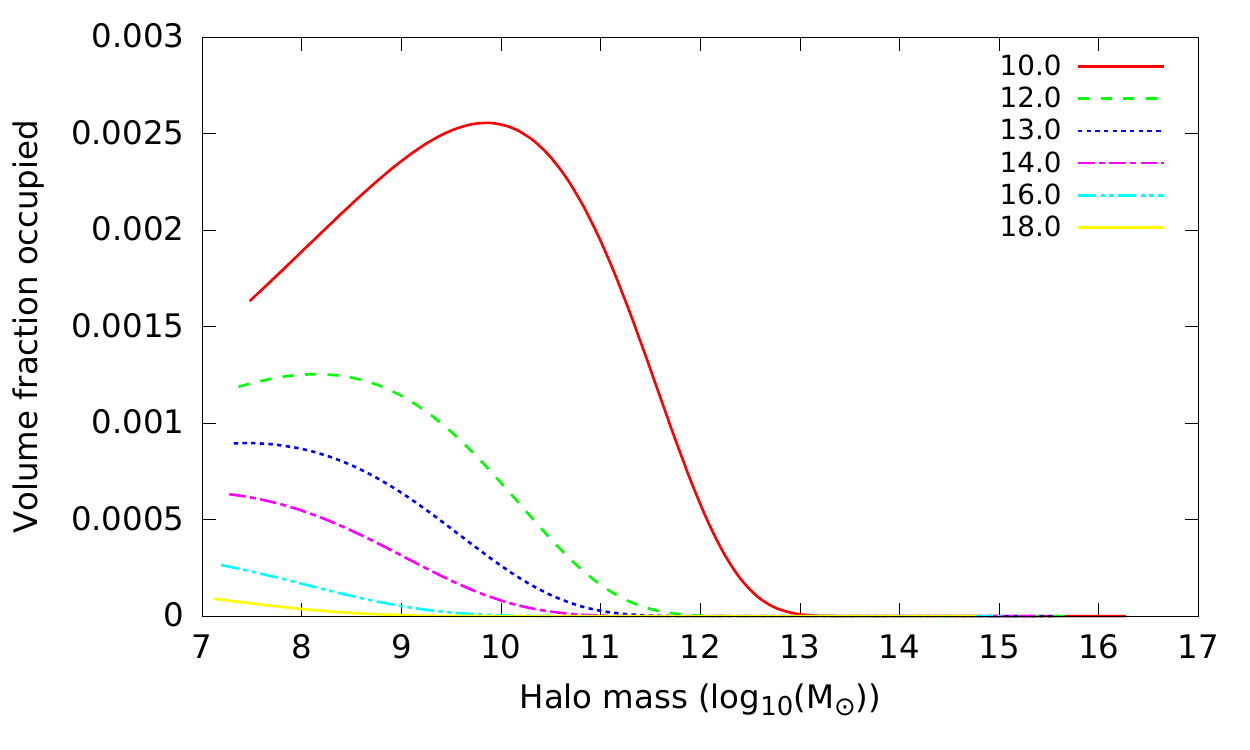}
		\caption{Evolution of the size distribution of self-ionized regions as a function of halo mass is shown for $\zeta = 10$.}
		\label{fig:ionbubbledis}
    \end{minipage}\hfill
\end{figure}

Figure~\ref{fig:ionzeta} shows the effect of $\zeta$ on the global ionization fraction $f_i$ of the universe. For higher value of $\zeta$, the reionization is completed at higher redshift. 
In Figure~\ref{fig:ionbubbledis} we show the 
distribution of volume fraction occupied by bubbles of different sizes and its evolution with redshift. We show the volume fraction as a function of halo mass, which can be related to the comoving size of  
the self-ionized bubble as: $R_x \simeq 0.09 \, {\rm Mpc} (M/10^8 M_\odot)^{1/3} \zeta^{1/3}$. Figure~\ref{fig:ionbubbledis} agrees with the results of \cite{FZH04a} (their Figure~2) for the set of parameters used by them. 
For the set of parameters used in this paper, the self-ionized bubbles 
are smaller, e.g. at $z = 12$ the peak of the bubble distribution corresponds
roughly to a scale of $R_x \simeq  10 \, \rm Mpc$ for \cite{FZH04a}, while it 
peaks at $R_x  \simeq 3 \, \rm Mpc$ for our case. 

\subsection{X-Ray Heating} \label{sec:xrayheat}

Photons of energy $E \gg 13.6 \, \rm eV$ are not absorbed in the \ion{H}{2} region
but escape into the surrounding medium. These photons ionize and heat the 
medium through  photoionization and secondary collisional ionization and excitation. These photons can ionize the medium to a level less than 10\% and can impart up to 20\% of their total energy
into heating the medium (e.g. \cite{1985ApJ...298..268S,Heating2001}). As the 
fraction of ionization in this process is generally tiny,  in our study,  we assume the medium outside the ionized region to be comprised of neutral hydrogen and neutral helium with primordial abundances. We hope to study the effect of partially ionized regions in future work.

A key fact in X-ray heating of the mostly neutral gas outside the \ion{H}{2} region is 
that photoionization cross section falls  as $1/\nu^3$ for energies far larger than the threshold energy.  As the low energy  X-ray photons are absorbed with higher probability, they contribute to heating the medium immediately surrounding the  \ion{H}{2} region  whereas the higher energy photons free-stream through the medium. As they are redshifted, their probability of absorption slightly increases. These photons uniformly heat up the whole IGM to some background temperature $T_{\text{bg}}$. 

In this section, we calculate the profile of  kinetic temperature  around self-ionized regions. In our analysis we separate the regions into near- and far-zones. In the near zone the heating is dominated 
by X-ray photons from an individual self-ionized region. In the far zone, 
the contribution from only the far away background sources is taken into account.  

\subsubsection{Near Zone}
To compute the temperature profile for an individual source at a redshift $z$,
we calculate the total energy absorbed (over the entire history of the source) by a point close to the source at redshift $z'$. In this subsection and the following one, primed quantities are calculated at the receiving point (point $P$), unprimed quantities are at the source (point $S$), and quantities with $0$ subscript are comoving quantities.

The energy attained by electrons due to ionization of species $i$ (here $i$ 
runs over hydrogen and neutral helium with their relative fractions $x_i=12/13$
and $1/13$ of the baryon number, respectively), per unit time, per unit comoving volume, per unit frequency (at $P$) is:
\begin{equation}
	 \frac{\mathrm{d}E_{\nu'}'(i)}{\mathrm{d}t'\mathrm{d}\nu'\mathrm{d}V_0} = (h\nu'-h\nu_i) \frac{\mathrm{d}N'_{\nu'}}{\mathrm{d}t'\mathrm{d}\nu'\mathrm{d}V_0} P(i,\nu')
\label{eq:xrayabso}
\end{equation}
Here $\mathrm{d}N'_{\nu'}/(\mathrm{d}t'\mathrm{d}\nu'\mathrm{d}V_0)$ is the number of X-ray photons (of frequency $\nu'$) received per unit time per unit
frequency per unit comoving volume and $P(i,\nu')$ gives the probability of 
ionization of species $i$  by a  photon of frequency $\nu'$ in a shell of thickness $dl'$:
\begin{equation}
	 P(i,\nu') = n_i' \sigma_i(\nu') \mathrm{d}l' 
\label{eq:probabso}
\end{equation}
Here $n_i' = x_i n_0 (1+z')^3$ is the local number density of species $i$, and  $\sigma_i (\nu')$ is the ionization  cross section of an atom by an X-ray photon of frequency $\nu'$ \footnote{We use here the approximate expressions for the frequency dependence of the ionization cross-section of hydrogen and 
neutral helium; for more precise formulas see e.g. \cite{1989agna.book.....O}. Our results are not affected by this choice.}:,
	\begin{align} 
        \sigma_i(\nu') = {\sigma_i}_0\left(\frac{\nu'}{\nu_i}\right)^{-3}, 
    \end{align}
with $\nu_i$ being the ionization threshold  of species $i$. 
We assume the  X-ray photon  luminosity to be given by a power law (e.g. \cite{21CMFAST} and references therein):
	\begin{equation}
        \dot{N}_\nu = \dot{N}_t \left(\frac{\nu}{\nu_\text{min}}\right)^{-\alpha}
        \label{eq:xraylum}
    \end{equation}
where $\nu_\text{min}$ is the lowest frequency (in the rest frame of the source) of X-ray photons escaping from  ionizing sources. 
We also have, 
	\begin{equation}
        \frac{\mathrm{d}\dot{N'}_{\nu'}}{\mathrm{d}\nu'} = \frac{\mathrm{d}\dot{N}_\nu}{\mathrm{d}\nu} 
\mathrm{e}^{-\tau(R_0,\nu')} 
		\label{eq:xrayrec}
    \end{equation}
The optical depth between two point separated by comoving distance $R_0$ is given by:
$\tau(R_0,\nu') = \int \sum_i P(i,\nu')$. 

We calculate the   X-ray luminosity of a self-ionized region following 
the prescription of the last section, which allows us to relate $\dot{N}_t$ to
$\zeta$ (Eq.~(\ref{eq:defzeta}))  and the comoving radius of the self-ionized region $R_x$:
	\begin{align}
		\dot{N}_t &= \frac{\text{Number of X-ray photons emitted}}{\text{time}} \nonumber \\
				&= \frac{\text{Number of X-rays emitted}}{\text{Number of Baryons in stars}} \frac{\text{Number of Baryons in stars}}{\text{Number of Collapsed Baryons}} \frac{\text{Number of Collapsed Baryons}}{\text{Time}} \nonumber \\
				&= N_{\text{heat}} f_\star \frac{\mathrm{d}(N_{\text{halo}} f_{\text{coll,ion}})}{\mathrm{d}t} 
	\end{align}
where $N_{\text{heat}}$  is the number of X-ray photons emitted per stellar baryons. $f_\star$ is the fraction of collapsed baryons that is converted into stars. $N_{\text{halo}} =4\pi/3 R_x^3 n_0$ is the number of baryons in a self-ionized region of radius $R_x$. $f_{\text{coll,ion}} = 1/\zeta$ is the collapsed fraction in an  ionized region.  We further assume that the collapsed fraction inside an ionization region follows the global collapsed fraction, $f_{\text{coll,g}}$, which we obtain from the excursion set formalism. This gives us:
	\begin{align}
         \frac{\mathrm{d}(N_{\text{halo}} f_{\text{coll,ion}})}{\mathrm{d}t} &= \frac{1}{\zeta}N_{\text{halo}}\frac{\dot{f}_{\text{coll,g}}}{f_{\text{coll,g}}} 
         \label{eq:collbar}
    \end{align}

Using Eqs.~(\ref{eq:xrayabso}) to~(\ref{eq:collbar}), using $dV_0 = 4\pi R_0^2 dR_0$, adding contribution due to all the species and integrating over all the frequencies  $\nu > \nu_\text{min}$, we get the energy
that goes into heating the medium per unit time per unit volume:
\begin{equation}
	\frac{\mathrm{d}E'_{\text{heat}}}{\mathrm{d}t'\;\mathrm{d}V_0} =  \frac{h f_H \alpha N_{\text{heat}} f_\star n_0^2 \nu_\text{min}^{\alpha}}{ 3 \zeta}\frac{ R_x^3}{R_0^2} \frac{(1+z')^{\alpha+3}}{(1+z)^{\alpha+1}} \frac{\dot{f}_{\text{coll,g}}}{f_{\text{coll,g}}} \int_{\nu_\text{min}'}^{\infty} {\nu'}^{-\alpha-4} \mathrm{e}^{-\tau(R_0,\nu')} \sum_i (\nu'-\nu_i)  x_i {\sigma_i}_0{\nu_i}^3 \mathrm{d}\nu' 
\label{eq:heatfin}
\end{equation}
Where, $\nu_\text{min}' = \nu_\text{min} (1+z')/(1+z)=$ the minimum frequency from the source that reaches P. We assume that $f_H =0.15$ is  the fraction of energy of emitted photoelectron that goes into heating the medium (\cite{1985ApJ...298..268S}, \cite{Heating2001}).

Eq.~(\ref{eq:heatfin}) gives the energy that goes  into heating the medium by X-ray photons  per unit time per unit comoving volume at distance $R_0$ from the center of a self-ionized  bubble of radius $R_x$.  To get the total increase in temperature due to this heating, we need to integrate this over time.  If $z'_c$ is the redshift at which we compute the heating profile, then to take into account the adiabatic cooling since higher redshift $z'$, we multiply by $(1+z'_c)^2/(1+z')^2$. We neglect all other cooling processes. $R_x(t')$, the radius of the given ionization region at time $t'$ in the past, is not a straightforward quantity to calculate as excursion set formalism doesn't give the time evolution of the radius of a particular self-ionized region. Given that the formalism allows us to compute the evolution of the average ionized fraction $f_i$, we assume  $R_x^3(t')=R_x^3(t)(f_i(t')/f_i(t))$. This gives,
	\begin{align}
		\frac{\mathrm{d}E'_{\text{heat}}}{\mathrm{d}V_0} &= \frac{h f_H \alpha N_{\text{heat}} f_\star n_0^2 \nu_\text{min}^{\alpha}}{ 3 \zeta} \frac{R_x^3}{R_0^2} (1+z'_c)^2 \nonumber \\
		& \quad \quad \int_{t(z_\star)}^{t(z_c)} \mathrm{d}t' \frac{f_i(t')}{f_i(t)} \frac{\dot{f}_{\text{coll,g}}}{f_{\text{coll,g}}} \left(\frac{1+z'}{1+z}\right)^{\alpha+1} \nonumber \\
		& \quad \quad \quad \int_{\nu_\text{min}'}^{\infty} \mathrm{d}\nu' {\nu'}^{-\alpha-4} \mathrm{e}^{-\tau(R_0,\nu')} \sum_i (\nu'-\nu_i)  x_i {\sigma_i}_0{\nu_i}^3 
	\end{align}
Finally, this allows us to compute the  increase in temperature due to 
a self-ionized region of radius $R_x$ at a distance $R_0$ from the center of the ionized region: 
	\begin{align}
		\Delta T' & = \frac{1}{n_0 k_B}\frac{\mathrm{d}E'_{\text{heat}}}{\mathrm{d}V_0} \nonumber \\
			&= \frac{h f_H \alpha N_{\text{heat}} f_\star n_0 \nu_\text{min}^{\alpha}}{3 k_B\zeta} \frac{ R_x^3}{R_0^2} (1+z'_c)^2 \nonumber \\
				& \quad \quad \int_{t(z_\star)}^{t(z_0)} \mathrm{d}t'\frac{f_i(t')}{f_i(t)} \frac{\dot{f}_{\text{coll,g}}}{f_{\text{coll,g}}} \left(\frac{1+z'}{1+z}\right)^{\alpha+1} \nonumber \\
				& \quad \quad \quad \int_{\nu_\text{min}'}^{\infty} \mathrm{d}\nu' {\nu'}^{-\alpha-4} \mathrm{e}^{-\tau(R_0,\nu')} \sum_i (\nu'-\nu_i)  x_i {\sigma_i}_0{\nu_i}^3 
\label{eq:fintemp}
\end{align}
We can estimate the typical energies of photons that are absorbed close to the source: at $z = 20$, for $\nu \simeq 100 \, \rm eV$, the photon is absorbed at a comoving distance $\simeq 3 \, \rm Mpc$ from the source while a photon of $1 \, \rm keV$
is absorbed at $\simeq 300 \, \rm Mpc$. Therefore, the low energy photons
are absorbed locally and determine the heating profile of the near zone 
while the high energy photons play the role of determining the evolution of the 
average temperature which we discuss next. One can compute the optical 
depth of high energy  photons to determine
the fraction of these  photons that are absorbed until  the epoch of 
interest; for $\nu= 2 \, \rm keV$, nearly 80\% of  the photons emitted at $z=20$ remain unabsorbed at $z=15$. This means that some of these photons are absorbed 
neither locally  nor do they participate in heating in the far zone; we return to this point in the next sub-section.  We also note that X-ray photons 
are not absorbed within the \ion{H}{2} region and hence the effective length for computing the optical depth is not $R_0$ but $R_0 - R_x$, which we take into account in our numerical computation. 

\subsubsection{Far Zone}
In the far zone, multiple sources contribute to the heating  at any point. 
As noted above, low frequency X-ray photons are preferentially absorbed 
close to the ionizing region in the near zone (Eq.~(\ref{eq:probabso})), while higher frequency photons escape far away from the ionizing  centers and contribute to 
global heating. This can be simplified when the distance traveled by the photon before being absorbed exceeds the mean distance between sources. As the comoving mean  distance between sources  is on the order of  $1 \, \rm Mpc$ at $z\simeq 20$ for many of the models we consider, it is  a good approximation for our study.  

In calculating far zone temperature, we essentially take into account all the X-ray frequencies emitted by all the sources since the time they were turned on. 
To calculate the increase in temperature due to all faraway sources, we choose a random point and calculate the increase in temperature at that point due to sources that lie in a shell of thickness $\mathrm{d}R_0$ at distance $R_0$ and then we integrate it over all $R_0$. We can take the upper limit of $R_0$ to correspond the redshift of star formation $R_\text{final}=R_0(z_\star)$. If we take the minimum value of $R_0$ to be 0, we get the total (average) temperature increase due to all sources over the history of the universe. 

Using the global ionization fraction for the redshift of the chosen shell $f_i(z(R_0))$, the volume of ionized gas inside this shell at comoving distance $R_0$ is, $4\pi R_0^2 \mathrm{d}R_0 f_i(z(R_0))$. Therefore, we can compute the impact of 
distant sources  by  replacing  $4\pi/3 R_x^3$ in Eq.~(\ref{eq:fintemp})  with $4\pi R_0^2 \mathrm{d}R_0 f_i(z(R_0))$ and integrate over $R_0$. This gives us: 
	\begin{align}
		\Delta T'_{\text{bg}} 
				= \frac{h f_H \alpha N_{\text{heat}} f_\star n_0 \nu_\text{min}^{\alpha}}{k_B\zeta} (1+z'_0)^2 & \int_{0}^{R_0(z_\star)}  \mathrm{d}R_0  \int_{t(z_\star)}^{t(z)} \mathrm{d}t'  \frac{\dot{f}_{\text{coll,g}}(t')}{f_{\text{coll,g}}(t')} f_i(t') \left(\frac{1+z'}{1+z}\right)^{\alpha+1} \nonumber\\
				&  \quad\quad \quad\quad \quad\quad\int_{\nu_\text{min}'}^{\infty} \mathrm{d}\nu' {\nu'}^{-\alpha-4} \mathrm{e}^{-\tau(R_0,\nu')} \sum_i (\nu'-\nu_i)  x_i {\sigma_i}_0{\nu_i}^3 
	\label{eq:globtemp}
	\end{align}

\subsubsection{Overlap}
During the initial phase of the evolution of the heated bubbles, when the ionized and partially heated fractions ($f_i$ and $f_h$, respectively) are small, the partially heated fraction grows and so does the background temperature. At low redshift, the heating bubbles will expand and start to overlap.
In this sub-section, we discuss  how we take  into account the impact of these overlaps.  Defining 
the  ionization volume fraction:
	\begin{align}
		f_i = \sum_{R_x}\frac{4\pi}{3}N(R_x)R_x^3
	\end{align}
where, $R_x$ are radii of ionization bubbles and $N(R_x)$ is the number density of bubbles with ionization radius $R_x$. Similarly, we can define the volume fraction due to heating bubbles to be,
	\begin{align}
		f_{hb} = \sum_{R_x}\frac{4\pi}{3}N(R_x)(R_h^3-R_x^3)
	\end{align}
This quantity can exceed unity due to significant overlaps as the universe evolves. Therefore, we define another quantity $f_h$, that corresponds to the  actual volume fraction occupied by heating bubbles. This can be derived by recognizing that, in the  case of overlap, within every heating bubble, there can be part of another heating bubble or ionization bubble: 
	\begin{align}
		f_h &= \sum_{R_x}\frac{4\pi}{3}N(R_x)(R_h^3-R_x^3) (1-f_i-f_h) \nonumber \\
			&= f_{hb} (1-f_i-f_h) \nonumber
	\end{align}
This gives us,
	\begin{align}
		f_h = \frac{f_{hb} (1-f_i)}{1+f_{hb}} \nonumber
	\end{align}
$f_h$ remains less than unity even if the value of $f_{hb}$ becomes much larger than unity and approaches $f_{hb}$ when the heating and ionization fractions are  small.  This allows us to successfully model overlap of heating bubbles. 

Also, since bubbles overlap, the temperature of a profile around an ionizing source should contain contribution due to other overlapping bubbles.
	\begin{align}
		T_p = T_s + T_o
	\end{align}
Here $T_p$,  the resultant  temperature, is the sum of $T_s$,  the  temperature  due  a nearby source and and  $T_o$, the average contribution due to overlaps. We recursively define $T_o$ as,
	\begin{align}
		T_o &= \sum_{R_x} \sum_s f_s (T_s+T_o) \nonumber \\
			& = \sum_{R_x} \sum_s \frac{4\pi}{3} N(R_x) \frac{f_h}{f_{hb}} ((R_s+\Delta R_s)^3 - R_s^3) T_s + f_h T_o \nonumber \\
			& = \frac{1}{1-f_h}\sum_{R_x} \sum_s \frac{4\pi}{3} N(R_x) \frac{f_h}{f_{hb}} ((R_s+\Delta R_s)^3 - R_s^3) T_s
	\end{align}
where, $R_x$ and $R_h$ are radii of chosen ionization bubble and outer radius of corresponding heating bubble respectively. $R_s$ and $\Delta R_s$ are inner radius of the shell with temperature $s$ around chosen bubble and the thickness of this shell respectively. We add $T_o$ to the fluctuations, but subtract $T_o/f_n$ from background to  maintain the energy budget. 

\subsubsection{Modelling}

\begin{figure}
    \centering
    \begin{minipage}{0.49\textwidth}
		\centering
		\includegraphics[width=1.0\textwidth]{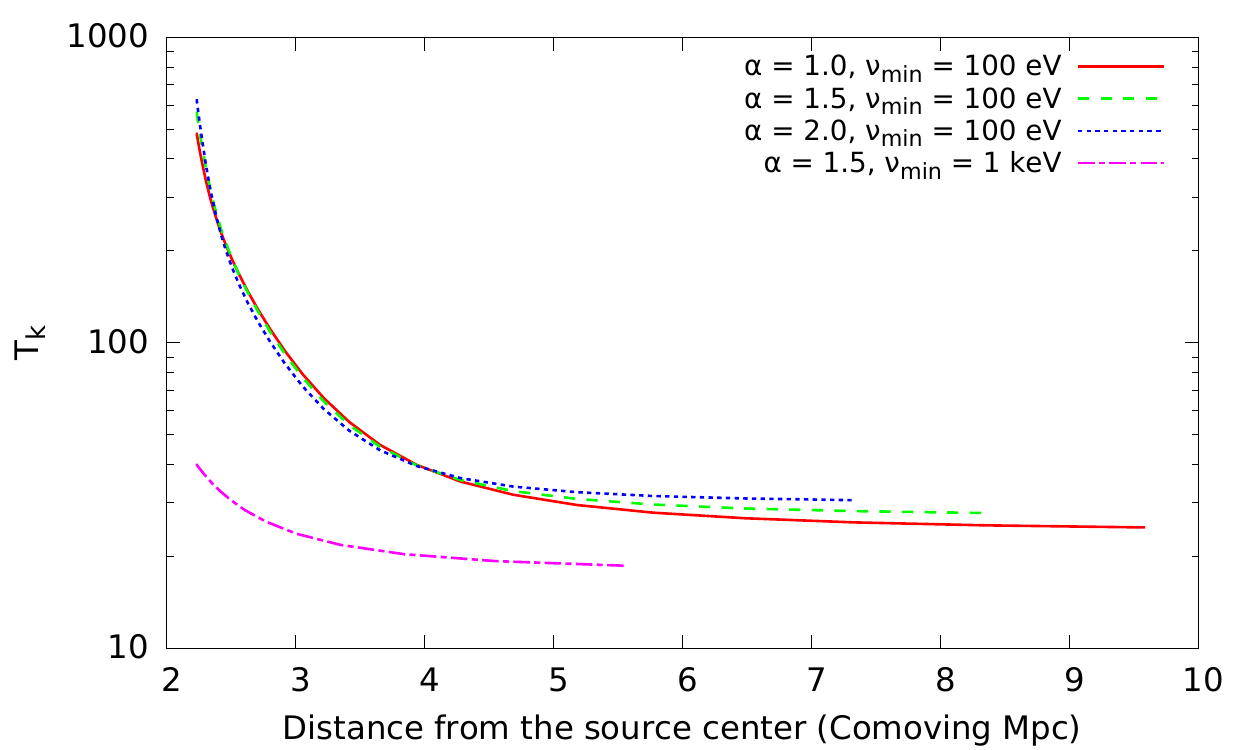}
		\caption{Heating profiles  around a bubble are shown at  $z=17$ for various  values of $\alpha$ and $\nu_\text{min}$ with $\zeta=10$ and $N_{\text{heat}}=1.0$. It is seen that for $\nu_{\rm min} > 1 \, \rm keV$ the temperature is 
smaller and the profile around a source is shallow or there is less  distinction between the near- and far-zone (for details see text).}
		\label{fig:heat_prof_17_10_1_100}
	\end{minipage}\hfill
	\begin{minipage}{0.49\textwidth}
		\centering
		\includegraphics[width=1.0\textwidth]{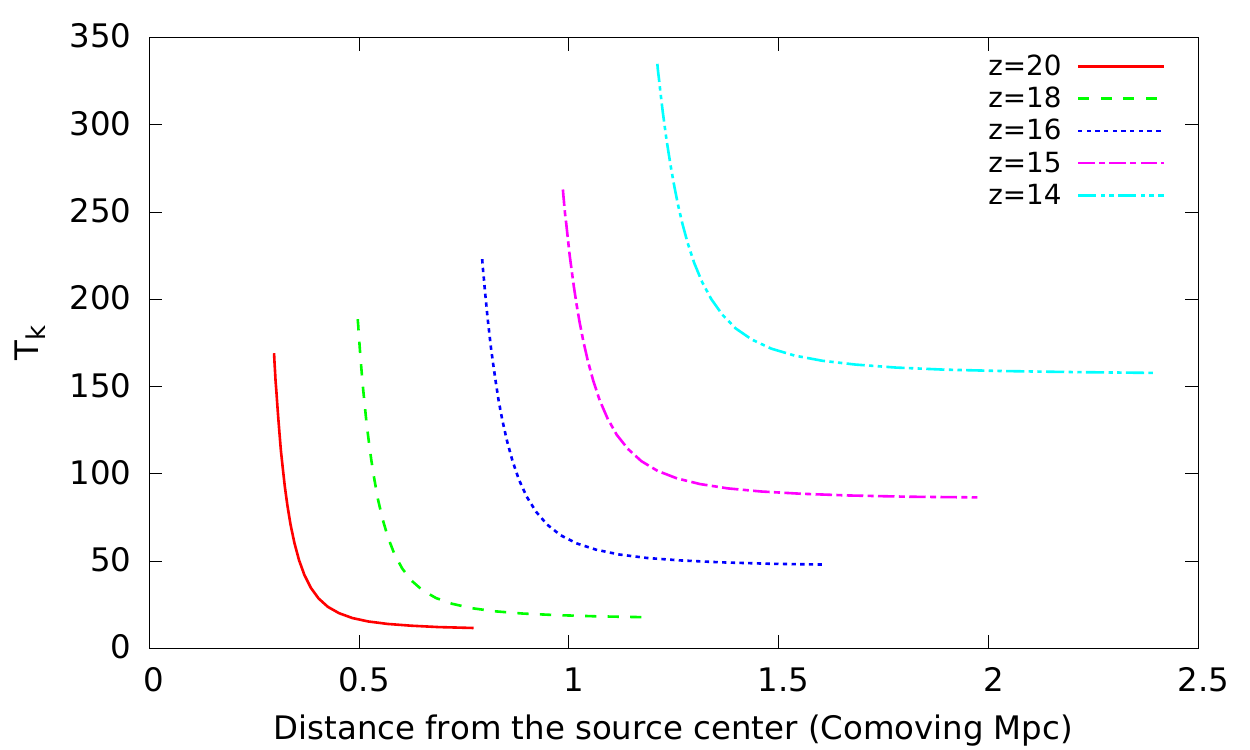}
		\caption{Evolution of the heating profile around an ionized  bubble is plotted for  $\alpha = 1.5$,  $\nu_\text{min} = 100 \, \rm eV$,  $\zeta=10$,  and $N_{\text{heat}}=1.0$. The size of the   fiducial ionized bubble is assumed to  grow as the mean ionized fraction in the universe. The smallest radius in each profile displayed is the size of the ionized region. The profiles shown reflect the growth of mean ionization fraction and  the increase in background temperature. }
		\label{fig:heat_prof_evol}
	\end{minipage}\hfill \\
	\begin{minipage}{0.49\textwidth}
		\centering
		\includegraphics[width=0.99\textwidth]{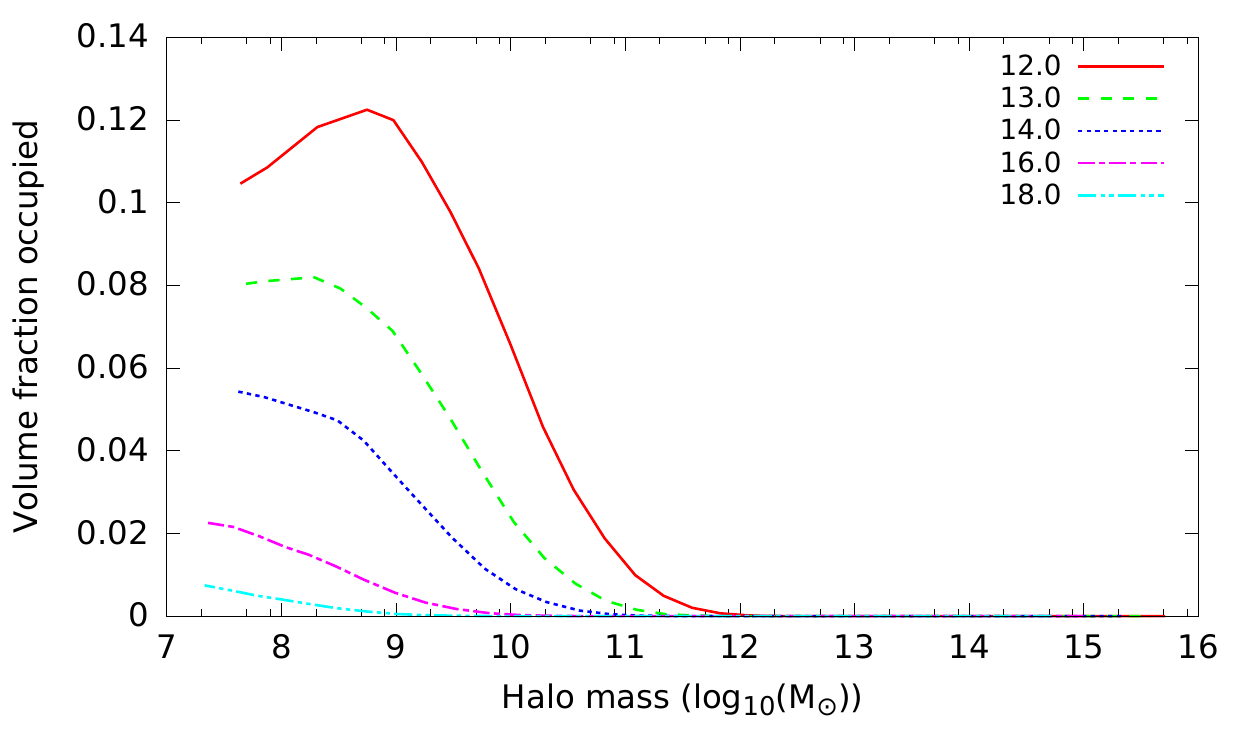}
		\caption{The evolution of the size  distribution of heated regions (see text for details)  as a function of halo mass is shown for $\zeta = 10$, $\alpha=1.5$ and $N_\text{heat}=1$. }
		\label{fig:heat_dist_1.5_10_1}
    \end{minipage}\hfill
\end{figure}

Eqs.~(\ref{eq:fintemp}) and~(\ref{eq:globtemp})  can be used to compute the 
temperature profile for a single self-ionized region and the evolution of global
heating. 

We use the following prescription for defining the near- and far-zones. The radius of the near zone is defined as the distance at which the temperature increase due to that source (over its history) is less than 1~K. 
The temperature in the far zone depends on the choice of minimum $R_0$ in Eq.~(\ref{eq:globtemp}). We choose  $R_0 = 0$ which means the photon energy absorbed in the near zone is also included. 
However, as we can independently estimate the total amount of energy
absorbed in the near zone, we subtract this energy from the energy budget 
used to estimate the far-zone temperature. The far zone temperature is then
added to the temperature profile of the near zone, to consistently take into 
account the fact that Eq.~(\ref{eq:globtemp}) gives the global temperature at all points including the near zone.

 We note  that while the 
definition of the radius of the locally heated region is somewhat arbitrary, its
impact on the break up of energy budget is negligible. Also, and 
importantly, the correlation functions as defined in the next sections are minimally affected by this  definition  as the correlation functions depend only upon the gradient of temperature.

We explore three parameters to model heating in this paper:
\begin{itemize}
  \item $\alpha =$ X-ray spectrum power index. We take three possible values, 1.0, 1.5 and 2.0, with the middle value to be standard case. For higher value of $\alpha$, there are more photons at low frequency. These photons more effectively heat up the medium since there is higher probability of them being absorbed. Therefore, with higher $\alpha$, the background temperature is high and the heating profiles are steeper.
  \item $N_\text{heat}$: Number of X-ray photons emitted per stellar baryon. For our study, we assume $N_{\rm heat}$ in the range: 0.1--10.0.
  \item $\nu_\text{min}$: Minimum X-ray frequency escaping the  source halo.  We take two possible values, 100~eV and 1~keV.  For a given $N_{\rm heat}$,  higher value of $\nu_\text{min}$ means the emitted  photons  are more energetic. They free stream into the medium and uniformly heat  medium with little fluctuations around source halos.
\end{itemize}

To explore more complicated models where the X-ray luminosity is not power law (e.g. \cite{LateHeat2}), we can take $N_\text{heat}$ and $\alpha$ to be a function of frequency $\nu$ and time. However, we do not explore such models in this paper.

In Figure~\ref{fig:heat_prof_17_10_1_100} we plot temperature profiles around an ionization bubble at  $z=17$ for  $\zeta=10$,  $N_{\text{heat}}=1.0$,  and three values of $\alpha$. The figure also displays a case when $\nu_{\rm min} = 1 \, \rm keV$. In this case, owing to the absence of low energy photons, the near-zone profile around the source is very shallow. Also, as noted above, the heating is suppressed in this case even in the far zone as many high energy photons remain unabsorbed (\cite{LateHeat2}). 
In Figure~\ref{fig:heat_prof_evol} we show the evolution of the heating profile around an ionized bubble (similar results have been obtained by e.g. \cite{2011MNRAS.417.2264V,2015MNRAS.447.1806G}). This figure captures the impact of the growth of ionized region and the average ionization fraction on the heating profile and the background temperature. In Figure~\ref{fig:heat_dist_1.5_10_1} we show the heating bubble size distribution corresponding to ionization bubbles from Figure~\ref{fig:ionbubbledis}.

\subsection{Lyman-$\alpha$ radiation} \label{sec:lyalpha}
As noted in section~\ref{sec:h1sig}, Lyman-$\alpha$ radiation plays an 
important role in determining the brightness temperature of \ion{H}{1} emission from
EoR. For EoR studies, all  the radiation between Lyman-$\alpha$ and Lyman-limit 
is referred to as Lyman-$\alpha$ and we shall follow this convention. Photons 
in this frequency band  are  not absorbed in the \ion{H}{2} region but  escape into the surrounding medium and redshift until its frequency nearly equals the resonant frequency of one of the  Lyman series lines.  Given the complicated frequency
structure of Lyman-series lines, these photons are absorbed at varying distances
from the source. Our aim here is to determine the conditions under which 
this radiation couples spin temperature of the \ion{H}{1} line $T_S$ to matter kinetic temperature $T_K$. This  coupling depends on two factors: the region of influence of the Lyman-$\alpha$ radiation and the coupling coefficient $y_\alpha$ (Eq.~(\ref{eq:tsbas})). 

First, we find the Lyman-$\alpha$ influence region, which is mainly determined  by the distance traveled by  the Lyman-$\beta$ photons to redshift to  Lyman-$\alpha$ frequency. If these photons were emitted at $z=z_e$ and absorbed at $z = z_a$ with 
$\nu_e = \nu_\beta$ and and $\nu_a = \nu_\alpha$, then, the comoving distance 
traveled by the photon before it is absorbed in an expanding universe is: 
	\begin{align}
		R_{\rm max} \simeq \frac{1422 \;\;\text{Mpc}}{(1+z_e)^{1/2}} \nonumber
	\end{align}
We note that  $R_{\rm max}$ is much larger  than the  mean distance between ionization bubbles at any redshift. For $\zeta=15$, the values of mean comoving distance between bubbles for redshift 25, 20, 15 is 9.66 Mpc, 2.61 Mpc and 1.08 Mpc respectively. Therefore Lyman-$\alpha$ regions are very large and merge very early. However,  this  would create  homogeneous coupling to \ion{H}{1} atoms   only if $y_\alpha$ is high enough (Eq.~(\ref{eq:tsbas})). 

Lyman-$\alpha$ coupling coefficient, $y_\alpha$ is a function of Lyman-$\alpha$ photon (physical) number density $n'_{\alpha}$ (\cite{Field1958,2004ApJ...602....1C}):
	\begin{align}
		y_\alpha = 5.9 \times 10^{11} \frac{n'_{\alpha}}{T_K^{3/2}}
\label{eq:lymancoup}
	\end{align}
For efficient coupling between kinetic temperature $T_K$ and spin temperature $T_S$ we need $y_\alpha \gtrsim T_\text{CMB}/T_K$.

We assume that Lyman-$\alpha$ contribution comes through two main factors: Lyman-$\alpha$ emitted from the sources and Lyman-$\alpha$ created due to X-ray photo-electrons (\cite{Heating2001}). The latter is generally negligible.  To calculate the number density of  Lyman-$\alpha$ photons from ionizing sources, we use the 
same method applied in the previous section, which seeks to express 
the Lyman-$\alpha$ photon luminosity in terms of the radii of ionizing 
regions. This method allows us to compute both the near- and far-zone 
contribution from Lyman-$\alpha$ photons. In this paper, we compute only
the far-zone contribution which comes from photons between Lyman-$\alpha$ and 
Lyman-$\beta$.  Assuming  flat spectrum between Lyman-$\alpha$ and Lyman-$\beta$, 
the number density of Lyman-$\alpha$ photons at a comoving distance 
$R_0$ from the source is:
	\begin{align}
		n'_{\alpha,\star} = \frac{\dot{N}_\alpha}{4\pi c R_0^2} \frac{2\Delta \nu_\alpha}{\nu_\beta-\nu_\alpha}\frac{(1+z')^3}{1+z} \nonumber
	\end{align}
Here $\dot{N}_\alpha$ is the Lyman-$\alpha$ luminosity of a given halo. $\Delta \nu_\alpha = \sqrt{8kT\ln(2)/m_p c^2} \nu_\alpha$ is the  Doppler line width and this  factor arises because at the source the photons are emitted with frequencies  between $\nu_\beta$ and $\nu_\alpha$, but the only frequencies which are absorbed at redshift $z'$ are in the range of $\Delta \nu_\alpha$ around $\nu_\alpha$.  

The Lyman-$\alpha$ luminosity,  $\dot{N}_\alpha$, can be expressed in
terms of the  size of ionization halo assuming that the Lyman-$\alpha$ luminosity
scales with ionizing luminosity with a factor $f_L$ and the balance between
ionization and recombination in the ionizing region:
	\begin{align}
		\dot{N}_\alpha = f_L\frac{4\pi}{3} n_0^2 \alpha_B C  R_x^3 (1+z)^3 \nonumber
	\end{align}

Here $10 < f_L < 100$ (e.g. \cite{2004ApJ...602....1C}). 

We can calculate the Lyman-$\alpha$ number density due to faraway sources in the same way as described in the previous section. In thin shell of width $\mathrm{d}R_0$ at comoving distance $R_0$ from the receiving point, the contributing ionization fraction is, $4\pi R_0^2 \mathrm{d}R_0 f_i(z)$. Therefore we integrate over $R_0$. Here we take the lower limit of the integral $R_0=0$ while the  upper limit is given  by the Lyman-$\alpha$ influence region.
	\begin{align}
		n'_{\alpha,\text{bg}} &= \frac{f_L n_0^2 \alpha_B C }{c} \int_0^{R_\text{max}}\mathrm{d}R_0 f_i(z) \frac{2\Delta \nu_\alpha}{\nu_\beta-\nu_\alpha}(1+z')^3(1+z)^2 \nonumber \\
			&= 4.71 \frac{ f_L n_0^2 \alpha_B C }{c^2} \sqrt{\frac{k}{m_p}} \frac{\nu_\alpha}{\nu_\beta-\nu_\alpha} (1+z')^3\int_0^{R_\text{max}}\mathrm{d}R_0 f_i(z) T_K(z)^{1/2} (1+z)^2 
\label{eq:lymanback}
	\end{align}
Here $R_\text{max}$ is distance corresponding to $z_\text{max} =(1+z')(\nu_\beta/\nu_\alpha) - 1$.

From Eq.~(\ref{eq:tsbas}) it follows that for $q \equiv y_{\alpha} T_{\rm K}/T_{\rm CMB} \ge 1$ 
we expect the spin temperature to relax to the matter temperature $T_K$. Using
Eqs.~(\ref{eq:lymancoup}) and~(\ref{eq:lymanback}) we can compute $y_\alpha$. 
Given the complicated temperature structure of the regions outside the ionized 
region, $q$ can vary substantially as it scales as $T_K^{-1/2}$. 
We find that for all the models we consider here, $q$ exceeds unity for $z < 20$, e.g. for  $\zeta=10$, $\alpha=1.5$, $N_{\rm heat}=0.5$, $f_L =100$, $C =2$ and $z=20$, the background temperature is 9.1~K, the value of $y_\alpha$ is 40 and $q = 6.3$. 

Here we do not calculate the effect of higher order Lyman transitions (eg. from Lyman-$\gamma$ to Lyman-$\beta$), since their total number density is less than photons between Lyman-$\beta$ and Lyman-$\alpha$. Moreover, they will be absorbed closer to the source.

\section{Auto-Correlation of Brightness Temperature $T_b$} \label{autoh1sig}

The auto-correlation of $\psi$ (Eq.~(\ref{psidef})) can be defined as: 
	\begin{eqnarray}
		\mu &=&\langle \psi_1\psi_2 \rangle-\langle \psi \rangle^2 \nonumber \\
			 &=& \langle n_1(1+\delta_1)(1-s_1)n_2(1+\delta_2)(1-s_2)\rangle-\langle n_1(1+\delta_1)(1-s_1)\rangle^2 \label{eq:defmu}
	\end{eqnarray}
Here, $n_1$, $\delta_1$ and $s_1$ are values of ionization, overdensity and heating ($T_\text{CMB}/T_S$) at point 1 (${\bf r_1}$). Similarly, $n_2$, $\delta_2$ and $s_2$ are values at point 2 (${\bf r_2}$).
It should be noted that the autocorrelation function $\mu$ is function of $r=|{\bf r_2}-{\bf r_1}|$ as the process of reionization is statistically homogeneous and isotropic. To calculate $\mu$, we need to find all the pairs of points which are separated by a distance $r$, and average them over the entire space. To compute the correlation function we  use geometric arguments to find the probability of pairs with given values and take their weighed average. 

Eq.~(\ref{eq:defmu}) can be greatly simplified if we assume that density has no correlation with ionization or heating ($\eta = \langle n \delta\rangle = 0$ and $\langle s \delta\rangle = 0$). In this work, we make the assumption that  cross-correlation between ionization and density as well as between heating and density is sub-dominant. We expect a positive correlation between ionization and density since dense regions collapse and get ionized first.  Using excursion set formalism  \cite{FZH04a} computed this cross-correlation  and  showed that it  is generally subdominant as compared to auto-correlation terms (section 3.5, Figure~5 of their paper). The correlation of heating with density is also expected to be positive as the dense regions surrounding ionization bubbles should have larger temperature.  However, on the scale of heated regions the density correlation is smaller at high redshifts. Simulations show that  the density-heating cross correlation  is sub-dominant (\cite{2015MNRAS.447.1806G}) as compared to other contributions.  
This gives us:
	\begin{align}
		\langle n_1n_2(1+\delta_1+\delta_2+\delta_1\delta_2)(1-s_1-s_2+s_1s_2)\rangle &= (1+\xi)\langle n_1n_2(1-s_1-s_2+s_1s_2)\rangle \nonumber
	\end{align}
Where $\xi=\langle \delta({\bf r_1})\delta({\bf r_2}) \rangle$  is the auto-correlation function of the \ion{H}{1} density perturbation; we compute $\xi$ using the $\Lambda$CDM model power spectrum, assuming the relative bias between the dark matter and the \ion{H}{1}, $b = 1$.  
And,
	\begin{align}
		\langle n_1(1+\delta_1)(1-s_1)\rangle &= \langle 1+\delta_1\rangle \langle n_1(1-s_1)\rangle \nonumber \\
            &= f_n -\langle n_1 s_1\rangle \nonumber
	\end{align}
Since $\langle \delta \rangle=0$ and $f_n=\langle n \rangle$ is defined as the dimensionless average neutral volume fraction at that redshift. This finally yields:
	\begin{align}
		\mu &= (1+\xi)(\langle n_1n_2\rangle-\langle n_1n_2s_1\rangle-\langle n_1n_2s_2\rangle+\langle n_1n_2s_1s_2\rangle)-(f_n -\langle n_1 s_1\rangle)^2 \nonumber
	\end{align}
We can greatly simplify correlation functions higher than second order. Let us first consider $\langle n_1n_2s_1\rangle$. This corresponds to the joint probability that the point ${\bf r} = {\bf r_1}$ is both neutral
and heated to a temperature such that $s = s_1$ while the second point is neutral. As noted above, $n$ can be either unity (neutral point) or zero (ionized point) while  $s$ can take any arbitrary value depending on the kinetic temperature. However, in our model $T_K \gg T_\text{CMB}$ if and only if that point is ionized ($T_K \sim 10^4 $ K). In other words, for a point $s=0$ if and only if that point has $n=0$. Therefore the condition of point 1 being neutral ($n_1=1$) is fulfilled by it being not heated to very high temperature ($s_1 \neq 0$).
This  allows us to simplify  higher point correlation functions as:
    \begin{align}
        \langle n_1n_2s_1\rangle &= \langle n_2s_1\rangle \nonumber \\
        \langle n_1n_2s_2\rangle &= \langle n_1s_2\rangle \nonumber \\
        \langle n_1n_2s_1s_2\rangle &= \langle s_1s_2\rangle \nonumber
    \end{align}
And since we choose two points randomly, we also have,
   	\[ \langle n_1s_2\rangle = \langle n_2s_1\rangle \]
Finally, we have,
	\begin{align}
		\mu &= (1 + \xi) (\langle n_1 n_2\rangle - 2\langle n_1 s_2\rangle +\langle s_1 s_2\rangle) - (f_n-\langle n_1 s_1\rangle)^2 
                \label{eq:corrfunexp}
	\end{align}
Here we have introduced cross-correlations between kinetic temperature (heating) and ionization ($\langle n_1 s_2\rangle$) and auto-correlation of heating ($\langle s_1 s_2\rangle$). These terms have significant effect in brightness temperature correlation. As noted above, all the 2-point correlations are function of distance between two points $|{\bf r_2} -{\bf r_1}|$.

In following subsections we explore certain simplifying cases and limits.

\subsection{Simplifying Cases}
\subsubsection{Uniform Heating} \label{sec:uniheatsec}
As a simplifying case, we examine a model where there is uniform heating outside ionized bubbles and  all the neutral gas of the IGM is at uniform temperature of $T_{\rm bg}$.
	\[ \psi = n(1+\delta)(1-s_b) \]
The correlation is,
	\begin{align}
		\mu &= (1-s_b)^2\langle n_1(1+\delta_1)n_2(1+\delta_2)\rangle-(1-s_b)^2\langle n_1(1+\delta_1)\rangle^2 \nonumber \\
			&= (1-s_b)^2((1+\xi)\langle n_1n_2\rangle- f_n^2) \label{eq:uniheat1}
	\end{align}
Here $s_b = T_{\rm CMB}/T_{\rm bg}$. At early times, $T_{\rm bg}$ can approach the adiabatically cooled temperature of the IGM gas which is smaller than $T_{\rm CMB}$.  If ionization fraction is too small, $\langle n_1 n_2\rangle \simeq f_n^2 \simeq 1$. This gives us,
	\[ \mu = \xi (1-s_b)^2 \]

At late times $T_{\rm bg} \gg T_{\rm CMB}$ owing to X-ray heating, driving $s_b$ to zero. Which gives,
	\begin{align}
		\mu = -f_n^2 + (1+\xi)\langle n_1 n_2\rangle 
	\end{align}
This result is consistent with \cite{ZFH04}.

\subsubsection{Correlation at Very Large and Small Scales}
We can compute the small and large scale limits  of the correlation function 
given by Eq.~(\ref{eq:corrfunexp}) under fairly general conditions. Both
ionization and heating  inhomogeneities  are caused by bubbles of a given 
size distribution, which determine the scales of correlation.  As discussed in the foregoing,   the size distribution evolves and 
bubbles could have complicated profiles. However, for scales  greater  than
the largest bubbles, the correlation function owing the ionization and heating 
inhomogeneities vanish, and the \ion{H}{1} correlation function is determined 
by only density perturbations. In this limits we get:
	\begin{alignat*}{3}
		\langle n_1 n_2\rangle &= \langle n_1\rangle \langle n_2\rangle &&= f_n^2 \nonumber \\
		\langle n_1 s_2\rangle &= \langle n_1\rangle \langle s_2\rangle &&= f_n \langle s\rangle \nonumber \\
		\langle s_1 s_2\rangle &= \langle s_1\rangle \langle s_2\rangle &&=\langle s\rangle^2 \nonumber
	\end{alignat*}
Therefore,
	\begin{align}
		\mu &= (1 + \xi) (\langle n_1 n_2 \rangle - 2\langle n_1 s_2\rangle +\langle s_1 s_2\rangle) - (f_n-\langle n_1 s_1\rangle)^2 \nonumber \\
			&= (1 + \xi) (f_n - \langle s\rangle)^2 - (f_n-\langle s\rangle)^2 \nonumber \\
			&= \xi (f_n - \langle s\rangle)^2 \label{eq:corrls}
	\end{align}
In this limit,  the correlation function scales as the density correlation function $\xi$. We also note that the correlation function vanishes when $f_n = \langle s \rangle$ (close to global heating transition). 

In the small scale limit (${\bf r_2} ={\bf r_1}$), we get:
	\begin{align}
		\mu &= (1 + \xi_0) (\langle n_1 n_1\rangle - 2\langle n_1 s_1\rangle +\langle s_1 s_1\rangle) - (f_n-\langle n_1 s_1\rangle)^2 \nonumber \\
			&= (1 + \xi_0) (f_n - 2\langle s\rangle +\langle s^2\rangle) - (f_n^2-2f_n\langle s\rangle+\langle s\rangle^2) \label{eq:corrss}
	\end{align}
Here $\xi_0 = \xi(0)$ or the correlation function computed at zero lag which equals RMS of density perturbations. Since Eq.~(\ref{eq:corrss})  gives an  RMS, it is always positive.  

We verify   Eqs.~(\ref{eq:corrls}) and~(\ref{eq:corrss}) as the large and small scale limits of the \ion{H}{1} correlation function computed using methods described in the next sections.

\begin{figure}
    \centering
    \begin{minipage}{0.45\textwidth}
		\centering
		\includegraphics[width=0.9\textwidth]{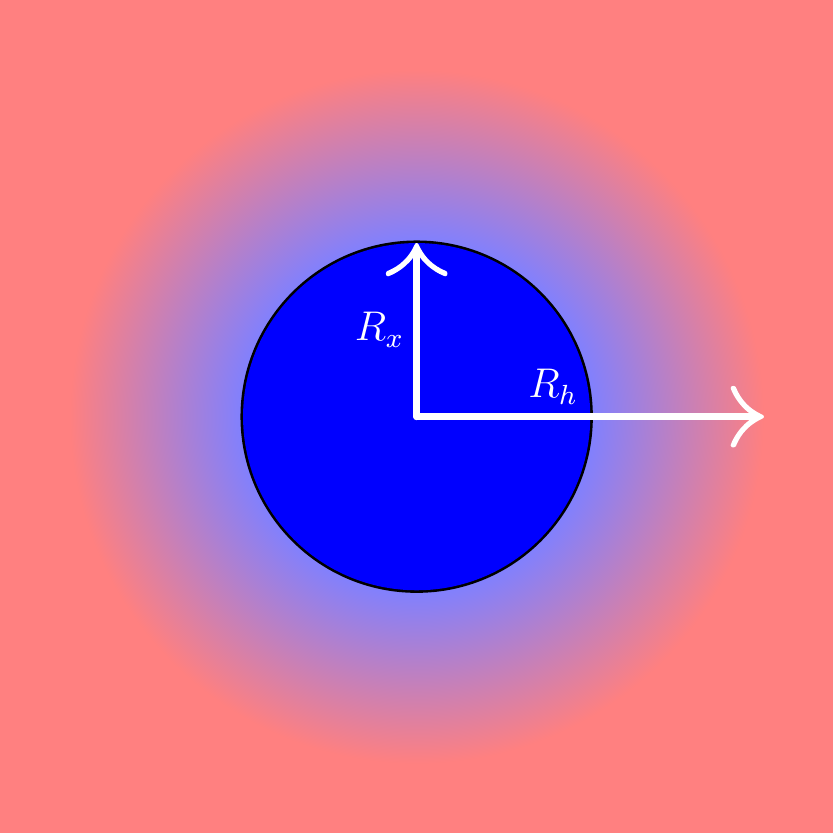}
		\caption{The topology of the ionized and the heated region typical of partially heated universe is shown. The ionized region of size $R_x$ (with sharp boundary) is seen to be surrounded by a heated fuzzy region of radius $R_h$. The colour scheme shows the temperature, $T$.   $T \simeq 10^4$ in the ionized regions and  it is much smaller in the heated region. It  falls as  distance from the source center and smoothly merges with the background. }
	\label{fig:PH}
    \end{minipage}\hfill
    \begin{minipage}{0.45\textwidth}
		\includegraphics[width=0.9\textwidth]{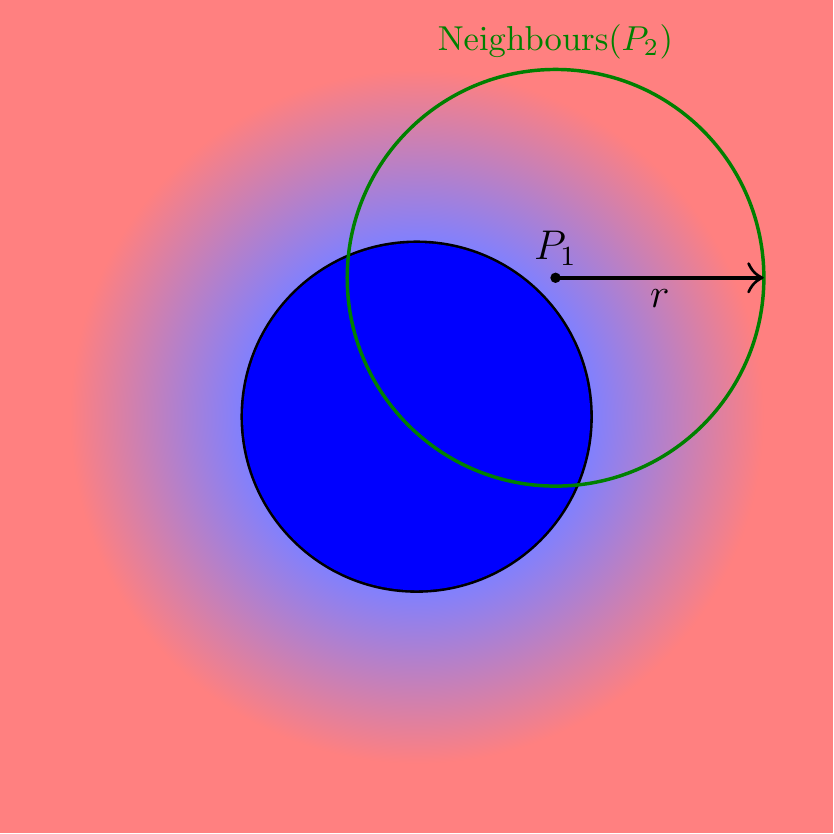}
		\caption{Randomly choosing a point and finding correlation with its neighbour at distance $r$. 
			\vspace{0.9in}}
		\label{fig:PH_corr}
    \end{minipage}
\end{figure}

\subsection{Modelling and Notations} \label{sec:model_and_notations}
Our aim in this paper is to analytically model the early phase of EoR. This issue can essentially be reduced to computing the auto-correlation of neutral fraction $\langle n_1n_2\rangle$, the ionization-heating cross-correlation $\langle s_1n_2 \rangle$, and the heating auto-correlation function $\langle s_1 s_2 \rangle$. We make several assumptions to make this problem analytically tractable. The main assumption is that  for a given self-ionized region   the heating and ionization bubbles are spherical and concentric and the ionization centers
are uncorrelated.  Each self-ionized and heating region can be treated as isolated  so long as both the ionization and heating fractions are small, which is expected during the early phase of reionization.  As discussed in section~\ref{sec:xrayheat}, X-ray heating can be split into near- and far-zone  effects. While the near-zone effects are owing to the vicinity of a self-ionized region, the far-zone effects take into account the impact of all the sources and their evolution. We model them both in this paper. An additional complication in the case of heating is that near-zone heating have a smooth profile around them, as opposed to ionized bubbles which have uniform ionization within the bubble with a sharp boundary. We explicitly account for the heating rate as a function of distance from the source and its smooth merger into the background. In Figure~\ref{fig:PH} we show the geometry of the self-ionized region and the heating zone beyond it. To compute correlations we assume two random points separated by a distance $r$ as shown in Figure~\ref{fig:PH_corr}.  The formalism used for the computation of  correlation functions is described in the Appendix. 

We assume that at a given $z$, the  ionized bubbles have various radii $R_x$. The number density of bubbles of radius $R_x$ is $N(R_x)$.  Between $R_x$ and $R_h$, the outer boundary of heating bubble (Figure~\ref{fig:PH}), we take shells of thickness $\Delta R(R_x, s)$, having nonzero temperature $s = T_{\rm CMB}/T_S$. A detailed description of notations  followed in this paper is as follows: 

\startlongtable
\begin{deluxetable}{c|cc}
	\tablecaption{Notations \label{tab:table}}
	\tablehead{
		\colhead{Symbols} & \colhead{Explanation} 
	}
	\startdata
	$\delta$ & Over-density of \ion{H}{1} gas \\
	$n$ & Ionization state of \ion{H}{1} gas: Neutral means $n=1$ and ionized means $n=0$ \\
	$s$ & Temperature state defined as $s=T_{\rm CMB}/T_S$ \\
	$\psi$ & Dimensionless brightness temperature: $\psi=n(1+\delta)(1-s)$ \\	
	$\xi$ & Auto-correlation of over density $\delta$: $\xi =\langle \delta_1\delta_2 \rangle$\\
	$\mu$ & Auto-correlation of dimensionless brightness temperature $\psi$: $\mu =\langle \psi_1\psi_2 \rangle-\langle \psi \rangle^2$\\
	$f_i$ & Average ionized volume fraction \\
	$f_n$ & Average neutral volume fraction \\
	$f_{hb}$ & Total volume fraction due to heating bubbles (without taking into account the overlaps)\\
	$f_h $ & Average heated volume fraction after taking into account the overlaps \\
	$f_b $ & Average background volume fraction \\
	$R_x$ & Radius of given ionization bubble\\
	$R_h$ & Outer radius of given heating bubble of, $R_h = R_h(R_x)$ \\
	$R_s$ & Inner radius of the shell with temperature $s$ around given bubble \\
	$\Delta R_s$ & Thickness of the shell with temperature $s$ around given bubble \\
	$N(R_x)$ & Number density of ionization bubbles of radius $R_x$ \\
	$P(n_p)$ & Probability of point $p$ being neutral \\
	$P(\widetilde{n_p})$ & Probability of point $p$ being ionized \\
	$P(\widetilde{n_p}(R_x))$ & Probability that point $p$ belongs to an ionization bubble of radius $R_x$ \\
	$P(s)$ & Probability that the given point has temperature $s$ \\
	$P(s_b)$ & Probability that the given point is in background \\
	$P(s, R_x)$ & Probability that the given point has temperature $s$ and lies in bubble of ionization radius $R_x$ \\
	\enddata
\end{deluxetable}
If the point is randomly chosen,
	\begin{align}
		P(\widetilde{n}(R_x)) &= N(R_x)\frac{4\pi}{3}R_x^3 \nonumber \\
		P(\widetilde{n}) &= \sum_{R_x} P(\widetilde{n_p}(R_x)) = \sum_{R_x} N(R_x)\frac{4\pi}{3}R_x^3 = f_i \nonumber \\
		P(n) &= 1- \sum_{R_x} N(R_x)\frac{4\pi}{3}R_x^3 = f_n\nonumber \\
		P(s, R_x) &= N(R_x)\frac{4\pi}{3} f_b ((R_s +\Delta R_s)^3-R_s^3) \nonumber \\
			P(s) &= \sum_{R_x} P(s, R_x) \nonumber 
	\end{align}
	The heated volume fraction can be expressed as,
	\begin{align}
		f_h &= \sum_{s} \sum_{R_x} P(s, R_x) \nonumber \\
			&= \sum_{R_x} N(R_x)\frac{4\pi}{3} f_b (R_h^3-R_x^3).
	\end{align}
	And the background volume fraction is,
	\begin{align}
		f_b & = \frac{1-f_i}{1+f_{hb}} \nonumber \\ 
			&= 1- f_i - f_h \nonumber \\
			&= 1- \sum_{R_x} N(R_x)\frac{4\pi}{3} (f_b (R_h^3-R_x^3) + R_x^3).
	\end{align}
	
	It should be noted that when $f_{hb}$  is small, $f_h$ approaches $f_{hb}$ and $f_b \simeq 1 - f_i-f_{hb}$, the values expected if the overlap is neglected. 
	
\subsection{Complete Model}
Our main aim is to  calculate ionization and heating correlations  for  epochs at which  ionization volume fraction is small. This ensures that ionization bubbles are separate and non-overlapping. However, as described in section~\ref{sec:xrayheat}, our formulation allows us to deal with  overlap of heating bubbles. We describe in detail our formalism to compute the neutral fraction auto-correlation, neutral fraction-heating cross-correlation and heating auto-correlation next. 

\subsubsection{Correlation of Neutral Region $(\langle n_1 n_2\rangle)$} \label{sec:n1n2sec}
We need to find the probability (fraction) of pairs with both points neutral (outside the ionization bubble).
	\begin{align}
		\langle n_1n_2\rangle &= 1^2\;P(n_1\cap n_2)+0\;P(n_1\cap \widetilde{n_2})+0\;P(\widetilde{n_1}\cap n_2)+0\;P(\widetilde{n_1}\cap \widetilde{n_2}) \nonumber \\
			&=P(n_1 \cap n_2) \nonumber
	\end{align}
Using Eqs.~(\ref{eq:AandB}),
	\begin{align}
		P(n_1 \cap n_2) &= P(n_1) - P(n_1\cap \widetilde{n_2}) \nonumber 
	\end{align}
First, we assume that point 2 is ionized. Therefore it lies in {\em some} ionized bubble. The statement that its neighbor (point 1) at distance $r$ lies in a neutral region means that point 1 lies outside {\em that} bubble and it lies outside {\em any other} bubble. Using Eq.~(\ref{eq:B3}),
	\begin{align}
		P(n_1 \cap \widetilde{n_2}) &= P((n_1(\text{out same}) \cap n_1 (\text{out other})) \cap \widetilde{n_2}) \nonumber \\
			&= P(n_1 (\text{out other}) |(n_1(\text{out same}) \cap \widetilde{n_2}))\;P(n_1(\text{out same}) \cap \widetilde{n_2}) \nonumber
	\end{align}
If we assume that the bubbles are uncorrelated and non-overlapping then $ P(n_1 (\text{out other}) |(n_1(\text{out same}) \cap \widetilde{n_2}))$  is the probability that point 1 is neutral given that point 2 lies in {\em some } bubble and point 1 lies outside {\em that } bubble. This quantity is equal to average neutral fraction $f_n$ in the present case. \footnote{If we assume that bubbles are uncorrelated and randomly distributed, then once a point is outside a certain bubble, its probability to be ionized or neutral is proportional to global ionized or neutral fractions respectively. However, this formalism applies because we are assuming infinite volume. If we had assumed finite volume (as would be the case for a simulation), the ionized volume fraction around an ionized bubble will be less than the global ionized volume fraction since we need to take into account the volume occupied by the said ionization bubble.
For a finite volume, this effect should cause   anti-correlation between bubbles. Throughout this paper, we assume infinite volume for averaging. \label{fn:uncorrelated_average}}

Now, we need to find $P(n_1(\text{out same}) \cap \widetilde{n_2})=$ Probability that point 2 is in an ionization bubble and point 1 lies out of that bubble. Point 2 can be in bubble with any radius $R_x$. Therefore,
	\begin{align} P(n_1(\text{out same}) \cap \widetilde{n_2}) &= \sum_{R_x}P(n_1(\text{out same})\cap \widetilde{n_2}(R_x)) \nonumber \\
		&= \sum_{R_x}P(\widetilde{n_2}(R_x))P(n_1(\text{out same})| \widetilde{n_2}(R_x))\nonumber \\
		&= \sum_{R_x}N(R_x)\frac{4\pi}{3}{R_x}^3 D(x,R_x) \nonumber
	\end{align}
Where $P(n_1(\text{out same})| \widetilde{n_2}(R_x))= D(x,R_x) =$ Probability that point 1 is out of the bubble given that point 2 is inside bubble of some radius $R_x$. 
Therefore,
	\begin{align}
		\langle n_1n_2\rangle &= f_n- f_n \sum_{R_x}N(R_x)\frac{4\pi}{3}{R_x}^3 D(r,R_x)   \label{eq:corrneu_case1}
	\end{align}
This expression  reduces to the  results  \cite{ZFH04} a single scale corresponding to size of ionized bubbles is taken for a fixed ionization fraction. 
It also follows from our discussion that the scenario envisaged in Figure~\ref{fig:PH} is valid at early time.  

\subsubsection{Correlation between Neutral region and Heating $(\langle n_1 s_2\rangle)$} \label{sec:n1s2sec}
We need to find the correlation between neutral points and points with $s \neq 0$.
	\begin{align}
		\langle n_1s_2\rangle &= s_b P(n_1 \cap s_b) + \sum_{0<s<s_b} s P(n_1 \cap s) \label{eq:n1s2def}
	\end{align}
Here $s_b$ corresponds to the background (far zone) temperature at any redshift. The first term can be written as:
	\begin{align}
		P(n_1 \cap s_b) &= P(s_b) - P(s_b\cap \widetilde{n_1}) \nonumber 
	\end{align}
We apply the procedure followed in the previous section.  As point 1 is ionized, it lies in some ionization bubble. The statement that its neighbor (point 2) at distance $r$ lies in background region means that point 2 lies outside the heating bubble corresponding to that ionization bubble and it lies outside any other heating bubble.
	\begin{align}
		P(s_b \cap \widetilde{n_1}) &= P(s_2 (\text{out other}) |(s_2(\text{out same}) \cap \widetilde{n_1}))\;P(s_2(\text{out same}) \cap \widetilde{n_1}) \nonumber
	\end{align}
As  the bubbles are assumed to be non-overlapping and uncorrelated $ P(s_b (\text{out other}) |(s_b(\text{out same}) \cap \widetilde{n_1}))$  gives the probability that point 2 is in background region given that point 1 lies in some  ionization bubble and point 2 lies outside heating bubble corresponding to that  ionization bubble. This probability equals the fraction of universe heated at background temperature, $f_b$ \footnote{Refer to footnote~\ref{fn:uncorrelated_average}.}.

Our next task is to compute  $P(s_2(\text{out same}) \cap \widetilde{n_1})$ which is the probability of point 1 being being in an ionization bubble and point 2 being out of the heating bubble corresponding to that ionization bubble. Given the distribution of radii of ionization bubbles, $R_x$, we have,
	\begin{align}
		P(s_2(\text{out same}) \cap \widetilde{n_1}) &= \sum_{R_x} N(R_x)\frac{4\pi}{3}{R_x}^3 P(s_2(\text{out same})| \widetilde{n_1}(R_x)) \nonumber
	\end{align}
We see that $P(s_2(\text{out same})| \widetilde{n_1}(R_x))= E(r,R_x,R_h(R_x))$, the  probability that point 2 is out of the heating bubble corresponding to the ionization bubble of radius $R_x$ in which point 1 lies.
Thus we have:
	\begin{align}
		P(n_1 \cap s_b) &= f_b-f_b\sum_{R_x} N(R_x)\frac{4\pi}{3}{R_x}^3 E(r,R_x,R_h(R_x)) 
	\end{align}
We also note that, in the limit $R_h \to R_x$, $E(r,R_x,R_h) \to D(r,R_x)$ which allows us to take the limit in which the region outside the ionizing bubbles is uniformly heated. 

Now, we need to find $P(n_1\cap s)$, where $0<s<s_b$. Here, point 2 is inside a heating bubble, but outside ionization bubble. Point 2 can be in heating bubble of any radius, thus,
	\begin{align}
		P(n_1\cap s)&= \sum_{R_x} P(s, R_x) P(n_1 | s(R_x))\nonumber
	\end{align}
Where $P(n_1 | s(R_x))$  is the probability that point 1 is in some neutral region given that point 2 is in the partially heated region of ionization bubble of size $R_x$ with $s_2=s$. If it is outside {\em that} ionization bubble, then its probability to be neutral = (1 - probability of there being an ionization bubble there). Using result of section~\ref{sec:n1n2sec},
	\begin{align}
		P(n_1 | s(R_x)) &= \left(1-\sum_{R'_x}N(R'_x)\frac{4\pi}{3}{R'_x}^3 D(r,R'_x)\right) C(r,R_s,R_s+\Delta R_s, R_x) \nonumber
	\end{align}
Where  $C(r,P,Q,R)$, as discussed in Appendix, gives the probability that if point 1 is between radii $P$ and $Q$ from the center of a sphere, then its neighbour point 2 at distance $r$ is outside radius $R$ of the same sphere. 
Therefore,
	\begin{align}
		P(n_1\cap s) &= \sum_{R_x} N(R_x) \frac{4\pi}{3} f_b ((R_s+\Delta R_s)^3-R_s^3) \left(1-\sum_{R'_x}N(R'_x)\frac{4\pi}{3}{R'_x}^3 D(r,R'_x)\right) C(r,R_s,R_s+\Delta R_s, R_x)
	\end{align}
This gives us the final expression: 
	\begin{align}
		\langle n_1s_2\rangle &= s_bf_b-s_bf_b\sum_{R_x} N(R_x)\frac{4\pi}{3}{R_x}^3 E(r,R_x,R_h) \nonumber \\
			& + \left(1-\sum_{R'_x}N(R'_x)\frac{4\pi}{3}{R'_x}^3 D(r,R'_x)\right) f_b \sum_{R_x}N(R_x)\frac{4\pi}{3} \sum_{s(R_x)} s\; ((R_s+\Delta R_s)^3-R^3_s) C(r,R_s,R_s+\Delta R_s, R_x) \label{eq:neus_cross1}
	\end{align}
In  writing Eq.~(\ref{eq:neus_cross1}) we have suppressed the dependence of $R_h$ and $R_s$ on $R_x$. 

We also need to calculate correlation between ionization and heating at the same point $(\langle n_1 s_1\rangle=\langle s\rangle$). The signal will be non-zero only if the chosen point is neutral. Therefore,
	\begin{align}
		\langle n_1 s_1\rangle &= s_b f_b + f_b\sum_{R_x} N(R_x)\frac{4\pi}{3} \sum_{s(R_x)} s\; ((R_s+\Delta R_s)^3-R^3_s)  \label{eq:neus_cross0}
	\end{align}
This quantity can also be represented as the global average of $s$, as $s=0$ where $n=0$.

\pagebreak

\subsubsection{Correlation of Heating $(\langle s_1 s_2\rangle)$} \label{sec:s1s2sec}
The signal will be non-zero only if neither of the points are completely heated or, equivalently, both the points lie outside ionized regions. Therefore, $s_1 \neq 0$ and $s_2 \neq 0$.
	\begin{align}
		\langle s_1 s_2\rangle &= s_b^2 P((s_1=s_b) \cap (s_2=s_b)) + 2 s_b \sum_{0<s<s_b} s P((s_1=s_b) \cap (s_2=s)) \nonumber \\
				&\quad +  \sum_{0<s_p<s_b} \sum_{0<s_q<s_b} s_p s_q P((s_1=s_p) \cap (s_2=s_q)) \label{eq:s1s2def}
	\end{align}
Following the logic of sections-\ref{sec:n1n2sec},
	\begin{align}
		P((s_1=s_b) \cap (s_2=s_b)) &= P(s_b) - P((s_1=s_b) \cap (s_2=\widetilde{s_b})) \nonumber 
	\end{align}
If point 2 is not in background region, it can be in ionized or heated profile region. In any case, its neighbour point 1 will be in background region only if it is outside the heated region of the bubble in which point 2 is.  Therefore, 
	\begin{align}
		P((s_1=s_b) \cap (s_2=s_b)) &= f_b - f_b \sum_{R_x}N(R_x)\frac{4\pi}{3}\left((R_h^3-R_x^3) f_bC(r,R_x,R_h,R_h)+R_x^3 E(r,R_x,R_h)\right) \label{eq:s1sb1}
	\end{align}
We have,
	\begin{align}
		P((s_1=s_b) \cap (s_2=s)) &= P( s_1(\text{out other})|(s_1(\text{out same})\cap(s_2=s))) P(s_1(\text{out same}) \cap (s_2=s)) \nonumber
	\end{align}
Where 
$P(s_1(\text{out other})|(s_1(\text{out same})\cap(s_2=s)))$ is the probability that point 1 is in background region given that {point 2 is partially heated with temperature $s$ and point 1 is not inside the bubble in which point 2 is}. This probability equals the fraction of the universe heated to background temperature, $f_b$ \footnote{Refer to footnote~\ref{fn:uncorrelated_average}.}.
$P(s_1(\text{out same})\cap (s_2=s))$  is the probability that point 1 is out of the bubble in which point 2 is, and point 2 is partially heated with temperature $s$. As point 2 can be in bubble with any ionization radius $R_x$, we have,
	\begin{align}
		P(s_1(\text{out same})\cap(s_2=s)) &= \sum_{R_x} N(R_x)\frac{4\pi}{3} f_b((R_s+\Delta R_s)^3-R^3_s)P(s_1(\text{out same})|(s_2=s)(R_x)) \nonumber
	\end{align}
$P(s_1(\text{out same})|(s_2=s)(R_x))=$ is the probability that point 1 is out of the bubble which has ionization radius $R_x$ and which contains point 2 with temperature $s$. This equals the probability that point 1 is out of the bubble with outer radius $R_h$ in which point 2 is located between radius $R_s$ and $R_s+\Delta R_s$. This gives us:
	\begin{equation}
		P((s_1=s_b) \cap (s_2=s)) = f_b \sum_{R_x} N(R_x)\frac{4\pi}{3}f_b ((R_s+\Delta R_s)^3-R^3_s)\;C(r,R_s,R_s+\Delta R_s,R_h)\label{eq:s1sb2}
	\end{equation}
In the case where both points are partially heated, these points can belong to the same bubble or they can belong to different bubbles, which gives:
	\begin{equation}
		P(s_1 \cap s_2) = P(s_1 \cap s_2 (\text{same}))+P(s_1 \cap s_2 (\text{diff})) \label{eq:s1s21}
	\end{equation}
Here $P(s_1 \cap s_2(\text{same}))$  is the probability that points 1 and 2 have temperature $s_1$ and $s_2$ respectively and they belong to the same bubble. We calculate $P(s_2(\text{same})|s_1(R_x))$, which is the probability that if point 1 is located in a bubble with ionization radius $R_x$ and has temperature $s_1$, then point 2 is in the same heating bubble with temperature $s_2$. If point 1 is located at distance between $R_{s_1}$ and $R_{s_1}+\Delta R_{s_1}$ from the center of the sphere, then fraction of its neighbours at distance $r$ which are outside the sphere of radius $R_{s_2}$ and inside sphere of radius $R_{s_2}+\Delta R_{s_2}$ can be computed. However, since bubbles can overlap, point 2 can be neutral or ionized, which leads to: 
	\begin{align}
		P(s_2(\text{same})|s_1(R_x)) &=( C(r,R_{s_1},R_{s_1}+\Delta R_{s_1},R_{s_2}) -C(r,R_{s_1},R_{s_1}+\Delta R_{s_1},R_{s_2}+\Delta R_{s_2}) )\nonumber \\
			& \quad\quad\left(1-\sum_{R'_x}N(R'_x)\frac{4\pi}{3}{R'_x}^3 D(r,R'_x)\right) \nonumber
	\end{align}
Therefore,
	\begin{align}
		\sum_{0<s_1<s_b} \sum_{0<s_2<s_b}\!\!\! s_1 s_2 P(s_1 \cap s_2 (\text{same})) &= \sum_{R_x} N(R_x) \frac{4\pi}{3} \sum_{s_1(R_x)} \sum_{s_2(R_x)}\!\!\! s_1 s_2\; f_b ((R_{s_1}+\Delta R_{s_1})^3-R_{s_1}^3) \nonumber \\
		& \quad (C(r,R_{s_1},R_{s_1}+\Delta R_{s_1},R_{s_2}) -C(r,R_{s_1},R_{s_1}+\Delta R_{s_1},R_{s_2}+\Delta R_{s_2})) \nonumber \\
			& \quad\quad\left(1-\sum_{R'_x}N(R'_x)\frac{4\pi}{3}{R'_x}^3 D(r,R'_x)\right) \label{eq:s1s2sa}
	\end{align}

Now we turn to  the second term on the RHS of Eq.~(\ref{eq:s1s21}). $P(s_1 \cap s_2 (\text{diff}))$ gives the  probability that point 1 has temperature $s_1$, point 2 has temperature $s_2$ and they both belong to different bubbles. Here we take a simple assumption that if point 2 is outside the bubble in which point 1 is, then its probability of having $s=s_2$ is equal to the global probability of $s_2$ temperature shell. Since point 1 and 2 can belong to bubbles of any size,
	\begin{align}
		P(s_1 \cap s_2 (\text{diff})) &= \sum_{{R_x}} P(s_1,{R_x}) C(r,R_{s_1},R_{s_1}+\Delta R_{s_1},{R_h})\sum_{{R'_x}} P(s_2,{R'_x})\nonumber
	\end{align}
which gives us,
	\begin{align}
		\sum_{0<s_1<s_b} \sum_{0<s_2<s_b}\!\!\! s_1 s_2P(s_1 \cap s_2 (\text{diff}))
			&= \sum_{{R_x}} N({R_x})  \frac{4\pi}{3}\sum_{s_1(R_x)}s_1 f_b ((R_{s_1}+\Delta R_{s_1})^3-R_{s_1}^3) \nonumber \\
			& \quad\quad C(r,R_{s_1},R_{s_1}+\Delta R_{s_1},{R_h}) \nonumber \\
			& \quad\sum_{{R'_x}} N({R'_x})\frac{4\pi}{3} \sum_{s_2({R'_x})} s_2 f_b ((R'_{s_2}+\Delta R'_{s_2})^3-{R'}_{s_2}^3)
			 \label{eq:s1s2dif}
	\end{align}
	
Using Eqs.~(\ref{eq:s1sb1}),~(\ref{eq:s1sb2}),~(\ref{eq:s1s2sa}), and~(\ref{eq:s1s2dif}), we  finally obtain the expression for the heating correlation function:
	\begin{align}
		\langle s_1 s_2\rangle&=s_b^2 f_b - s_b^2 f_b\sum_{R_x}N(R_x)\frac{4\pi}{3}\left(f_b (R_h^3-R_x^3) C(r,R_x,R_h,R_h)+R_x^3 E(r,R_x,R_h)\right)\nonumber \\
			&\; + f_b\sum_{{R_x}} N({R_x})\frac{4\pi}{3}\sum_{s(R_x)}s ((R_{s}+\Delta R_{s})^3-R_{s}^3) C(r,R_{s},R_{s}+\Delta R_{s},{R_h}) \nonumber \\
			& \quad \left(2 s_b f_b + f_b \sum_{{R'_x}} N({R'_x})\frac{4\pi}{3} \sum_{s'({R'_x})} s' ((R'_{s'}+\Delta R'_{s'})^3-{R'}_{s'}^3)\right) \nonumber \\
			&\; + f_b\sum_{R_x} N(R_x) \frac{4\pi}{3} \sum_{s_1(R_x)} \sum_{s_2(R_x)}\!\!\! s_1 s_2\; ((R_{s_1}+\Delta R_{s_1})^3-R^3_{s_1}) \nonumber \\
			& \quad\times (C(r,R_{s_1},R_{s_1}+\Delta R_{s_1},R_{s_2}) -C(r,R_{s_1},R_{s_1}+\Delta R_{s_1},R_{s_2}+\Delta R_{s_2})) \nonumber \\
			& \quad\quad\left(1-\sum_{R'_x}N(R'_x)\frac{4\pi}{3}{R'_x}^3 D(r,R'_x)\right)
			 \label{eq:s1s2fin}
	\end{align}

We also calculate the  correlation of heating at the same point $(\langle s_1 s_1\rangle = \langle s^2\rangle)$. The signal will be non-zero only if the chosen point is not heated to very high temperature. Therefore,
	\begin{align}
		\langle s_1 s_1\rangle &= s_b^2 f_b + f_b \sum_{R_x} N(R_x)\frac{4\pi}{3} \sum_{s(R_x)} s^2\; ((R_s+\Delta R_s)^3-R^3_s) \nonumber
	\end{align}
This quantity is global average of $s^2$.

\pagebreak

\subsection{A Simple Model: Uniform heating} \label{sec:simple_uni}
In this case we assume a limit where there is no heating profile around ionization bubble. We should get back the result derived in subsection \ref{sec:uniheatsec}.

If there are no heating shells, the second term of Eq.~(\ref{eq:n1s2def}) and second and third terms of Eq.~(\ref{eq:s1s2def}) can be dropped. In this limit we also obtain, $R_h = R_x$, $E(r,R_x,R_x)=D(r,R_x)$ and $f_b = f_n$.

Simplifying the results of section~\ref{sec:n1n2sec}, \ref{sec:n1s2sec} and \ref{sec:s1s2sec}:
	\begin{align}
		\langle n_1n_2\rangle &= f_n - f_n \sum_{{R_x}}  N({R_x}) \frac{4\pi}{3} {R_x}^3 D(r,{R_x}) \nonumber \\
		\langle n_1s_2\rangle &= s_bf_n-s_bf_n\sum_{R_x} N(R_x)\frac{4\pi}{3}{R_x}^3 D(r,R_x) \nonumber \\
		\langle s_1 s_2\rangle &=s_b^2 f_n - s_b^2 f_n\sum_{R_x}N(R_x)\frac{4\pi}{3}R_x^3 D(r, R_x)\nonumber \\
		&\langle n_1 s_1\rangle = s_b f_b = s_b f_n  \nonumber
	\end{align}

The total correlation is given by:
	\begin{align}
		\mu &= (1 + \xi) (\langle n_1 n_2\rangle - 2\langle n_1 s_2\rangle +\langle s_1 s_2\rangle) - (f_n-\langle n_1 s_1\rangle)^2 \nonumber \\
		&= (1 + \xi) (1-2s_b+s_b^2) f_n \left(1-\sum_{{R_x}}  N({R_x}) \frac{4\pi}{3} {R_x}^3 D(r,{R_x})\right) - (f_n- s_b f_n)^2 \nonumber \\
		&=  (1-s_b)^2 ((1 + \xi)\langle n_1n_2\rangle - f_n^2) \nonumber 
	\end{align}
This expression agrees with Eq.~(\ref{eq:uniheat1}) from section~\ref{sec:uniheatsec}. Our   model goes to correct limit for this simplified case. 

Here if we take only one bubble size, we have,
	\begin{align}
		\mu &=  (1-s_b)^2 f_n ((1 + \xi)(1-(1-f_n) D(r,{R_x})) - f_n) \label{eq:uniform_flat_onesize}
	\end{align}

\subsection{A Simple Model: One bubble size, flat heating profile} \label{sec:flatpro}
One of the principle aims of this paper is to establish the new scales that emerge in \ion{H}{1} correlation function for a  partially heated universe. For fully heated universe ($T_S \gg T_{\rm CMB}$ in neutral regions), these scales are determined by the size distribution of ionized regions. In the partially heated case, there is a separation between unheated and heated neutral regions. This situation is expected to introduce news scales linked to  the size of heated regions. Our detailed analysis of the physical conditions that exist during the early phase of reionization are difficult to interpret in terms of demarcated ionized, heated, and unheated regions, because unlike ionized regions which have sharp boundaries, the heated regions have shallow profiles which smoothly merge into the background. However, for the purposes of understanding our formalism  we consider a simple model:  a single bubble size (both ionized and heating) and the heating profile with uniform temperature (flat). Thus there are small ionization bubbles embedded in larger heated bubbles. We first ignore density fluctuations for simplicity in this section and later present the results including these perturbations. In this case, there are only three values of $\psi=n(1-s)$ present in the universe: $\psi_i$, $\psi_h$, $\psi_b$.  $\psi_i = 0$ since $n=0$ inside the ionized region. In the heated and background regions respectively, 
	\begin{align} 
		\psi_h &= (1-s_h) = 1-\frac{T_{\text{CMB}}}{T_{\text{heat}}} \nonumber \\
		\psi_b &= (1-s_\text{bg}) = 1-\frac{T_{\text{CMB}}}{T_{\text{bg}}} \nonumber
	\end{align}
If $N$ is the number density of bubbles then the total ionized volume fraction of the universe and the heated volume fraction without correcting for overlap are respectively,
	\begin{align} 
		f_i &= \frac{4\pi}{3} R_x^3 N \nonumber \\
		f_{hb} &= \frac{4\pi}{3} (R_h^3-R_x^3) N \nonumber
	\end{align}
where $R_x$ is the ionization bubble radius and $R_h$ is the heating bubble radius.
Allowing for overlaps, the actual heated volume fraction and the remaining non-heated volume fraction are respectively:
	\begin{align} 
		f_h &= \frac{f_{hb}(1-f_i)}{1+f_{hb}} \nonumber \\
		f_b &= 1 - f_i -f_h \nonumber
	\end{align}
This allows us to calculate the  dimensionless temperature correlation:
	\begin{align} \mu &= \langle \psi_1 \psi_2 \rangle - \langle \psi \rangle^2 \nonumber \\
		&= \psi_h^2 P ((\psi_1=\psi_h) \cap (\psi_2=\psi_h)) + 2\psi_h\psi_b P((\psi_1=\psi_h) \cap (\psi_2=\psi_b)) + \psi_b^2 P((\psi_1=\psi_b) \cap (\psi_2=\psi_b)) - \langle\psi\rangle^2 \nonumber
	\end{align}
Now we derive each term separately,
	\begin{align}
		P((\psi_1=\psi_h) \cap (\psi_2=\psi_b)) &= P(\psi_1=\psi_h) P((\psi_2=\psi_b)|(\psi_1=\psi_h)) \nonumber \\
			&= f_h f_bC(r,R_x,R_h,R_h) \label{flaprof2}
	\end{align}
The other terms give:
	\begin{align}
		P ((\psi_1=\psi_h) \cap (\psi_2=\psi_h)) &= P(\psi_1=\psi_h) -P((\psi_2 \neq \psi_h) \cap (\psi_1=\psi_h)) \nonumber \\
			&= f_h - P((\psi_2=\psi_b) \cap (\psi_1=\psi_h)) - P((\psi_2=\psi_i) \cap (\psi_1=\psi_h)) \nonumber \\
			&= f_h - f_hf_b C(r, R_x, R_h, R_h) - f_if_h C(r,0,R_x,R_h) \nonumber \\
			& \quad \quad- f_i(1-f_i)(C(r,0,R_x,R_x)-C(r,0,R_x,R_h)) \label{flaprof1}
	\end{align}
In Eq.~(\ref{flaprof1}), the third and fourth terms represent the  correlation of ionized and heated regions of  different bubbles and the  same bubble, respectively.
	\begin{align}
		P((\psi_1=\psi_b) \cap (\psi_2=\psi_b)) &= P(\psi_1=\psi_b) - P((\psi_1=\psi_b) \cap (\psi_2 = \psi_h)) - P((\psi_1=\psi_b) \cap (\psi_2  \psi_i)) \nonumber \\
			&= f_b - f_bf_h C(r,R_x,R_h,R_h) - f_bf_i C(r,0,R_x,R_h) \label{flaprof3}
	\end{align}
Putting the terms in  Eqs.~(\ref{flaprof1}),~(\ref{flaprof2})~and(\ref{flaprof3}) together, we have:
	\begin{align}
        \mu &= \psi_h^2 (f_h - f_hf_b C(r, R_x, R_h, R_h) - f_if_h C(r,0,R_x,R_h) \nonumber \\
			& \quad \quad- f_i(1-f_i)(C(r,0,R_x,R_x)-C(r,0,R_x,R_h))) \nonumber \\
				&\quad + 2  \psi_h\psi_b f_h f_bC(r,R_x,R_h,R_h) \nonumber \\
				&\quad + \psi_b^2 f_b(1 - f_h C(r,R_x,R_h,R_h) - f_i C(r,0,R_x,R_h) ) \nonumber \\
				&\quad - (f_h\psi_h+f_b\psi_b)^2 \label{eq:flatpro_wden}
	\end{align}
If the impact of density perturbations is included, we get:
	\begin{align}
		\mu		&= (1+\xi) \left[\psi_h^2 (f_h - f_hf_b C(r, R_x, R_h, R_h) - f_if_h C(r,0,R_x,R_h) \right.\nonumber \\
			& \quad \quad- f_i(1-f_i)(C(r,0,R_x,R_x)-C(r,0,R_x,R_h))) \nonumber \\
				&\quad + 2  \psi_h\psi_b f_h f_bC(r,R_x,R_h,R_h) \nonumber \\
				&\quad + \psi_b^2 f_b(1 - f_h C(r,R_x,R_h,R_h) - f_i C(r,0,R_x,R_h) ) \left. \right] \nonumber\\
				&\quad-(\psi_b f_b + \psi_h f_h)^2 \label{eq:flatpro_fin}
	\end{align}
To verify the validity of our formalism, we need to consider Eqs.~(\ref{eq:flatpro_wden}) and~(\ref{eq:flatpro_fin}) in different limits: (a) at large scales, all the functions $C(.,.,.,.)$ tend to unity. In this case, Eq.~(\ref{eq:flatpro_wden}) vanishes and Eq.~(\ref{eq:flatpro_fin}) approaches the correct
large scale limit (Eq.~(\ref{eq:corrls})), (b) $\psi_b = \psi_h$. In this case there is no distinction between the heated bubble and the background and we expect the correlation information from heated bubbles to disappear. In this case Eq.~(\ref{eq:flatpro_fin}) reduces to  Eq.~(\ref{eq:uniform_flat_onesize}), the case in  which only ionized bubbles and density perturbations contribute to the correlation.  One sub-case of this scenario is when both $\psi_b$ and $\psi_h$ approach unity, the limit in which the entire universe is uniformly heated at high temperatures $T_S \gg T_{\rm CMB}$, (c) finally, if we assume: $R_h \to R_x$, $f_h=0$, $f_b=f_n$, we get:
	\begin{align}
		\mu	 &= (1+\xi) \left[\psi_b^2 f_b(1-f_i C(r,0,R_x, R_h))\right] -(\psi_b f_b)^2 \nonumber \\
		 		&= (1+\xi) \left[(1-s_b)^2 f_n(1-(1-f_n) C(r,0,R_x, R_x))\right] -((1-s_b) f_n)^2 \nonumber \\
		 		&= (1-s_b)^2 f_n ((1+\xi)(1-(1-f_n) D(r,R_x))- f_n) \nonumber 
	\end{align}
This is the same result as Eq.~(\ref{eq:uniform_flat_onesize}). 

The agreement of our formulation with the expected results in different 
limits shows we have taken into account the relevant physical processes 
 in our study. 

{\it negative correlation}: Our formalism allows for negative correlation. 
Such a situation might arise if $s >1$ inside the heated bubble and $s <1$ 
in the background.  However, we do not find many instances of negative 
correlation in all the cases we study here even when we neglect density 
correlations. As $\xi$ is positive for all the scales we consider in this 
paper, inclusion of this term ensures we do not get negative correlation in 
any case. We check that we  can generate negative correlation  by assuming  the centers of 
ionizing centers to be anti-correlated.

\section{Results} \label{sec:res}

The brightness temperature correlation is caused by density, 
ionization, and heating inhomogeneities. The main aim of this paper is to study
the era dominated by heating inhomogeneities. There are two main effects 
in modelling the correlations in this era: (a) the near-zone
effect that introduces new scales corresponding to heated  
bubbles  around self-ionized bubbles (Figures~\ref{fig:heat_prof_17_10_1_100} and~\ref{fig:heat_prof_evol}). The correlation function  during the partially heated era is determined by the scales of  these bubbles which are much larger than the ionization bubbles. (b) the evolution of $s = T_{\rm CMB}/T_K$ in the far zone starting from  an era where $s$ can exceed unity.

We explore four modeling parameters in this paper: photoionization efficiency $\zeta$, X-ray spectral index $\alpha$, number of X-ray photons per stellar baryon, $N_{\rm heat}$, and minimum X-ray frequency escaping the source halo $\nu_{\rm min}$. These parameter have already been introduced in sections~\ref{sec:photion} and~\ref{sec:xrayheat}. $\zeta$ is constrained by Planck results that fix the optical depth to the reionization surface; $\zeta$ in the range 10--15 is in agreement with these results  (Figure~\ref{fig:ionzeta}) (\cite{Planck2016}).  We take runs in the redshift range 10--20.

\subsection{Simple model}

\begin{figure}
	\centering
	\begin{minipage}{0.49\textwidth}
		\centering
		\includegraphics[width=0.99\textwidth]{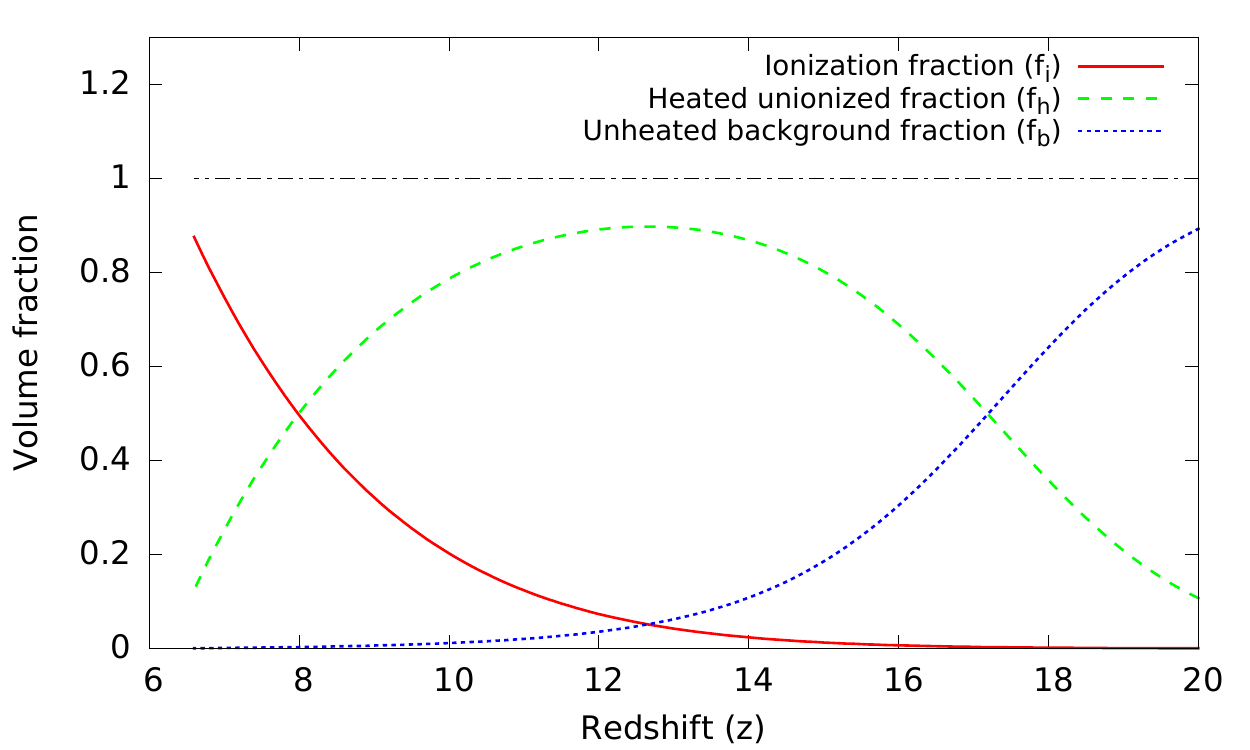}
		\caption{Evolution of ionized and heating fractions is shown for a fiducial model (described in the text)  for $\zeta = 10$, $N_{\rm heat} = 1$ and $\alpha = 1.5$}
	\label{fig:frac_all}
	\end{minipage}\hfill
	\begin{minipage}{0.49\textwidth}
		\includegraphics[width=0.99\textwidth]{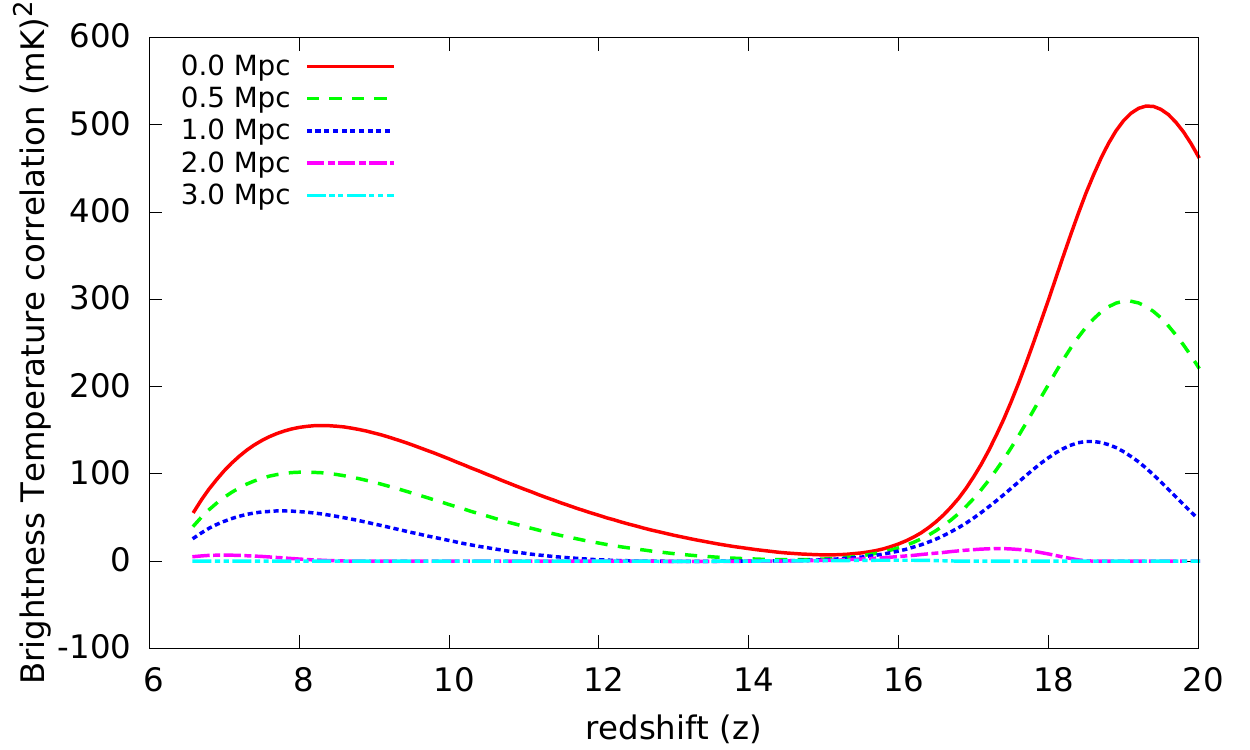}
		\caption{The evolution of correlation function is shown for a set of scales for the model  in Figure~\ref{fig:frac_all}.}
		\label{fig:flat_pro}
	\end{minipage} \hfill \\
	\begin{minipage}{0.49\textwidth}
		\centering
		\includegraphics[width=0.99\textwidth]{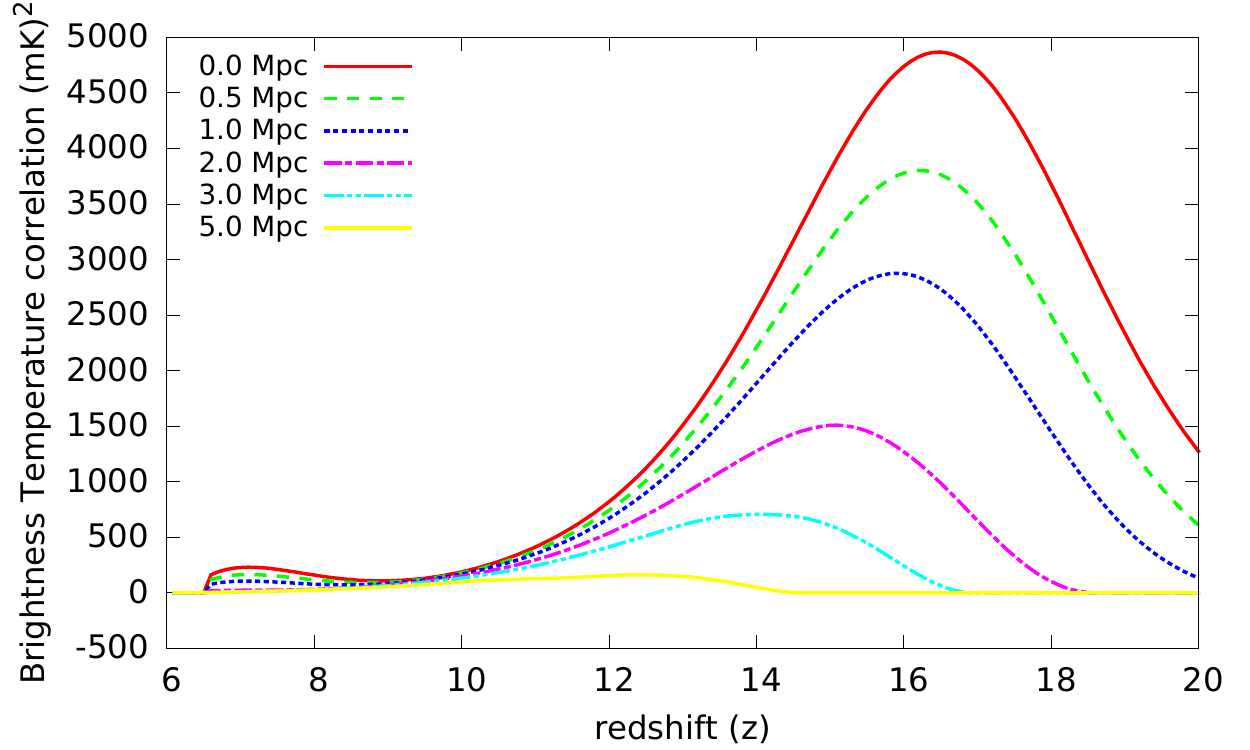}
		\caption{The evolution of correlation function  is displayed for a set of scales for a model in which the background temperature is held constant (see text for details).}
		\label{fig:flat_pro_merge}
    \end{minipage}\hfill
\end{figure}

To understand the evolution of correlation function at a given scale, we 
first consider the simple model based on an initial bubble distribution 
given by an ionized bubble of a single size which is surrounded by a heating
bubble of a single profile temperature (flat heating profile) with the ratio of heating to ionized bubble size
remaining constant (section~\ref{sec:flatpro}). We also neglect the impact 
of density perturbations in this case. We show the evolution of ionized, 
heated, and background  fractions---$f_i$,  $f_h$ and $f_b$---for such a case 
in Figure~\ref{fig:frac_all}. The initial radius of the ionized (heated) bubble is assumed 
to be $0.1 \, \rm Mpc$ ($0.7 \, \rm Mpc$). The initial  ratio
of heated and ionized fraction is the cube of the ratio of these radii. Initially nearly 90\% of the universe is in
the phase outside the heated bubbles. As the universe evolves, the ionized 
bubbles grow and so do the heated bubbles, resulting in an increase in both 
the ionized and heated fraction with a decrement of background fraction. This 
process is accompanied with an increase in the background and bubble temperature. At certain redshift, the heated bubble begin to merge driving the background fraction  to zero.  Eventually, the ionized fraction becomes large enough to drive the heated fraction to zero. 

The evolution of correlation function (normalized using Eq.~(\ref{overallnorm})) for a set of scales is shown in Figure~\ref{fig:flat_pro} \footnote{We also 
show the evolution of the RMS which corresponds to the  plot for 
$r = 0$ to guide the eye. For an experiment, the relevant quantity would 
be the RMS smoothed with  the three-dimensional resolution of the 
radio interferometer, which, as discussed below could be around 3--5~Mpc for 
ongoing and upcoming experiments. Therefore, the measured RMS would always be
smaller than the quantity shown in Figure~\ref{fig:flat_pro}.}. Here the background temperature is assumed to evolve according Eq.~(\ref{eq:globtemp}) for modeling parameters: $\zeta=10$, $N_{\rm heat}=1.0$ and $\alpha=1.5$, while the heating bubble temperature is kept at a constant value above the background temperature. Initially the correlation function is small which is expected because  the function tends to zero 
as $f_i$ and $f_h$ approach zero in the absence of density perturbations. The correlation function rises as $f_h$ increases and then decreases again owing to multiple reasons. 

There are 
three distinct reasons that can  wipe  out information on the correlation
scales generated by  heating inhomogeneities: (a)  increase in temperature
in the heated bubbles and in the background:  when these temperatures rise substantially
above $T_{\rm CMB}$, $s = T_{\rm CMB}/T_S$ is driven to zero causing  both autocorrelation of $s$
and its cross-correlation with ionization inhomogeneities to approach zero. 
This is the primary cause of the evolution of the correlation function as 
seen in Figure~\ref{fig:flat_pro}. 
(b) a decrease in the gradient of temperature between the heated bubble and 
the background.  This effect  plays an important role when fuzzy boundaries of 
heating regions in taken into account, which we discuss in the next section. 
It can also be achieved when $\nu_{\rm min}$ 
is increased as seen in Figure~\ref{fig:heat_prof_17_10_1_100}. In this case, the heating inside 
the bubble decreases and most of the X-ray photons are used in raising the 
background temperature. We discuss this case in the next section, (c) merging 
of bubbles: this process destroys the distinction between  heated bubble
and background thereby erasing correlation information on the scales of the 
bubbles. The difference between this case and case (b) is that the latter 
is possible  for even small heating fractions, $f_h$. 

 All these reasons play some role in determining the transition
from heating to ionization inhomogeneities regime. For the parameters we consider in this paper, the 
effect of both (a) and (c)  can be suppressed by considering a small $N_{\rm heat}$ while the scenario considered in (b) can be achieved by varying $\nu_{\rm min}$.

This behavior is generic to all models even though the  evolution on 
range of scales  displayed in Figure~\ref{fig:flat_pro} could change as it is determined 
by the sizes of heating bubbles. For instance, in Figure~\ref{fig:flat_pro} the correlation on
the scales of $\simeq 3 \, \rm Mpc$ remains close to zero at all times owing to our choices of initial scales.  The magnitude of correlation function is also determined by the background and heated  temperatures. Both rise as the universe evolves, decreasing $s$ and therefore decreasing the correlation function. In Figure~\ref{fig:flat_pro}, $T_h^{-1} f_h + T_b^{-1} f_b$ reaches $T_{\rm CMB}^{-1}$ at $z \simeq 16$ ; at this redshift
the global \ion{H}{1} signal vanishes and the universe makes a transition from 
being observable in the \ion{H}{1} signal from absorption to emission.  However, 
as discussed above, the redshift at which the correlation 
function reaches its minimum is determined by a multitude of other 
causes and this transition is reached at $z \simeq 15$.  

The minimum of correlation function around $z \simeq 15$ signals  the beginning 
of the phase in which the universe is uniformly heated. The signal at this 
time reaches nearly zero for all scales in our case because $f_i \simeq 0.01$ at the time of heating transition and therefore ionization inhomogeneities are small and we 
ignore density perturbations. As ionized fraction increases, the ionization 
inhomogeneities start rising, reaching a peak around $f_i \simeq 0.5$ and subsequently decline as the universe becomes fully ionized. It should be noted 
that the peak of the correlation function when it is dominated by ionization 
inhomogeneities is smaller than when it is determined by heating inhomogeneities. This is expected as $s$ is larger than unity in the earlier phase and it 
is zero during the latter phase. 

To isolate the impact of merging from the effect of heating on the 
evolution of correlation function, 
we show in Figure~\ref{fig:flat_pro_merge} a different  model which is also 
based on the evolution of ionization and heating fractions shown Figure~\ref{fig:frac_all}. In this case, the 
initial background temperature is $5 \, \rm K$ and the temperature inside the 
heated bubble is $10 \, \rm K$ and the temperatures are kept at their initial value throughout. Therefore, in this case, the heating inhomogeneities are destroyed by merging of heating bubbles and not heating, which delays the transition to uniform heating regime as compared
to Figure~\ref{fig:flat_pro}. 

Figures~\ref{fig:flat_pro} and~\ref{fig:flat_pro_merge} allow us to identify the relevant physical processes involved 
in the modelling of the correlation function in the phase when heating 
inhomogeneities dominate, the end of this phase (owing to either heating 
above CMB temperature or merger), and the transition to ionization inhomogeneities domination phase. As we shall notice later,  the features seen in the Figures are  also  present when more exact modelling 
is attempted. In this paper, we assume the centers of ionization bubbles to be uncorrelated. If the centers are correlated, new correlation scales can emerge,  we briefly discuss this scenario later in appendix \ref{sec:Correlations}.

\subsection{Complete model}

\begin{figure}
    \centering
    \begin{minipage}{0.49\textwidth}
		\centering
		\includegraphics[width=1.0\textwidth]{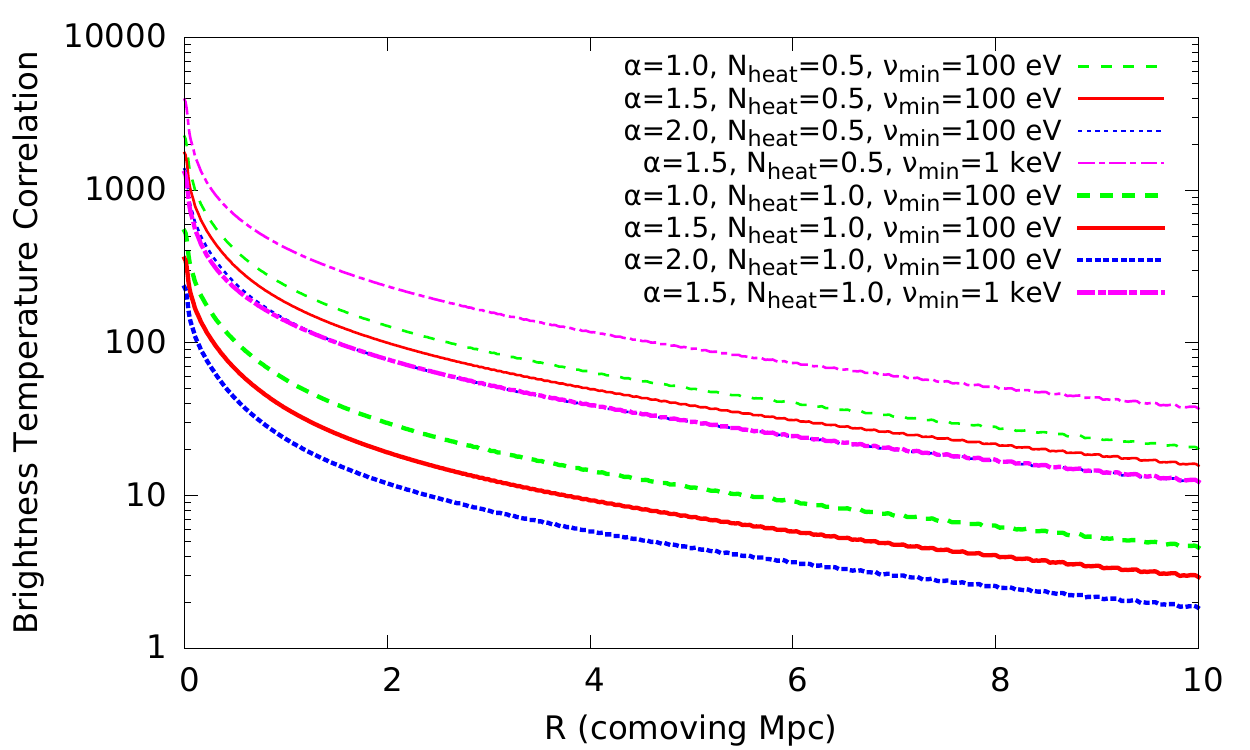}
		\caption{Two-point  correlation functions  are displayed for  different  values of $N_{\text{heat}}$, $\alpha$, and $\nu_{\rm min}$  for  $\zeta=10$ at $z = 17$}
		\label{fig:corr_comp_z17}
    \end{minipage}\hfill
    \begin{minipage}{0.49\textwidth}
		\centering
		\includegraphics[width=1.0\textwidth]{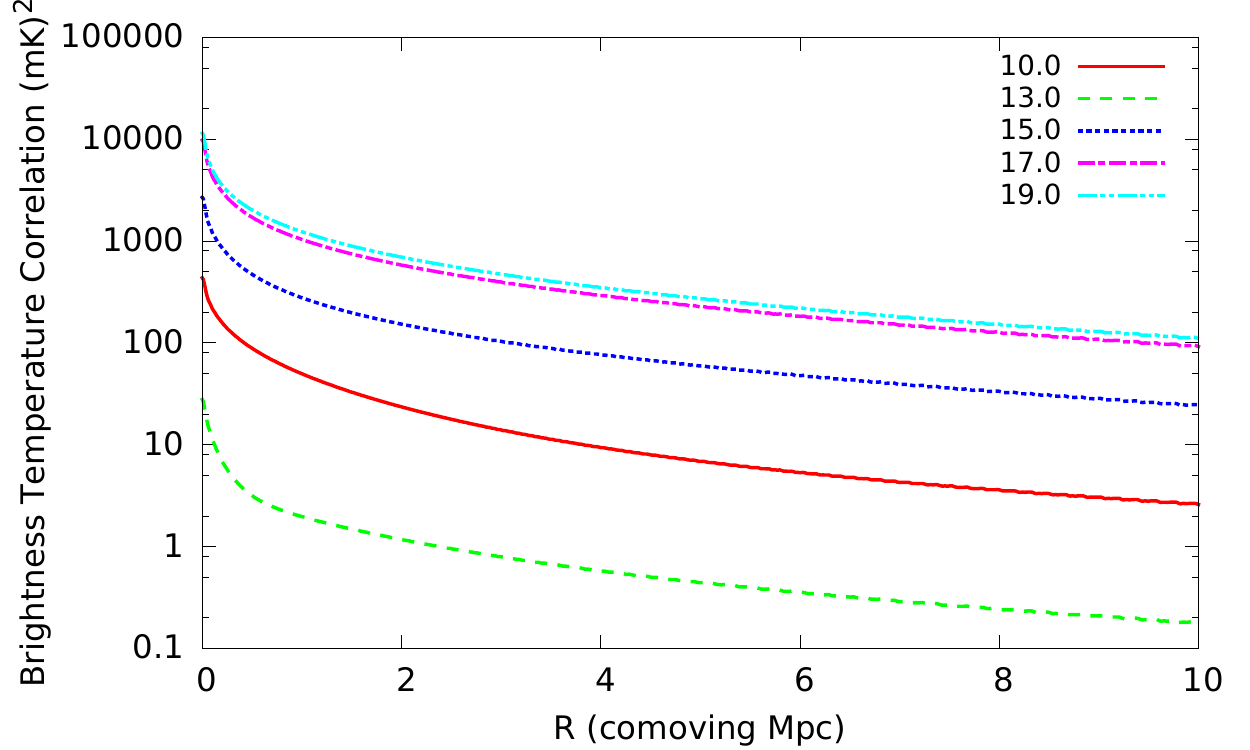}
		\caption{The evolution of  correlation function for $\alpha=1.5$, $\zeta=10$, $N_{\text{heat}}=0.1$ is displayed.}
		\label{fig:correvol_01}
    \end{minipage}\hfill \\
    \begin{minipage}{0.49\textwidth}
		\centering
		\includegraphics[width=1.0\textwidth]{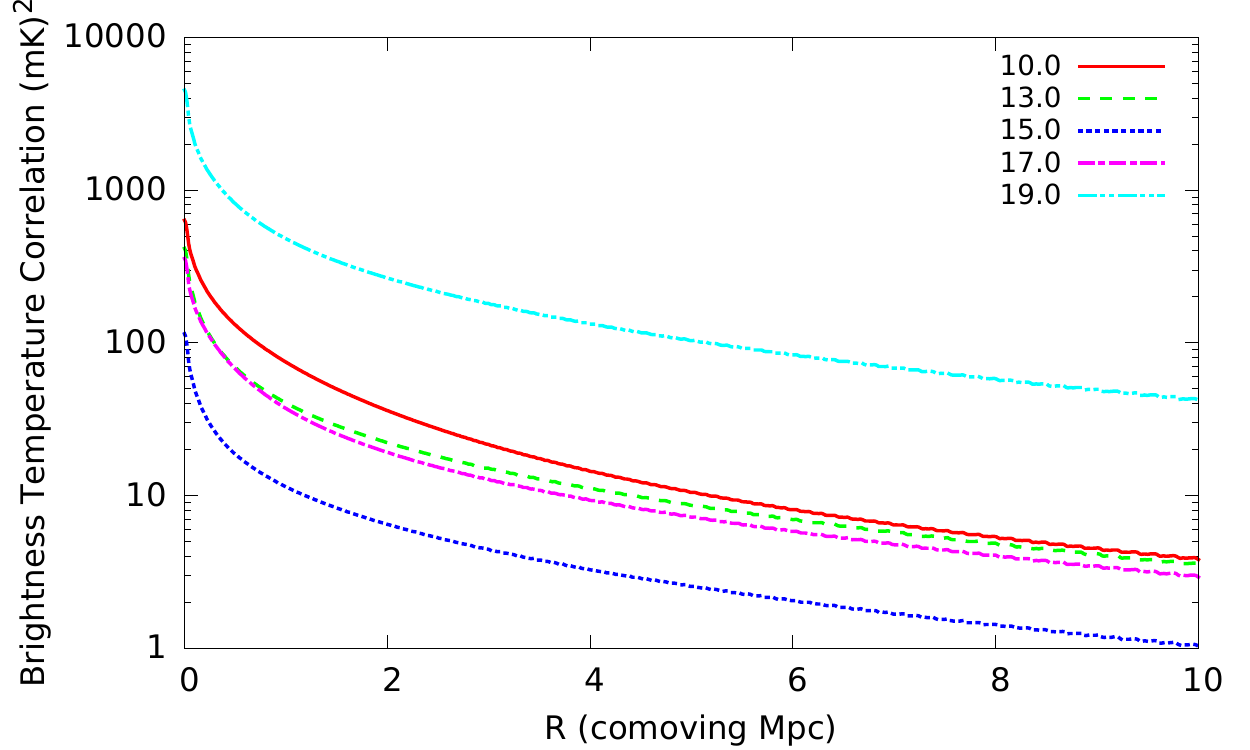}
		\caption{The evolution of  correlation function for $\alpha=1.5$, $\zeta=10$, $N_{\text{heat}}=1$ is displayed. \\}
		\label{fig:correvol_10}
    \end{minipage}\hfill
        \begin{minipage}{0.49\textwidth}
		\centering
		\includegraphics[width=1.00\textwidth]{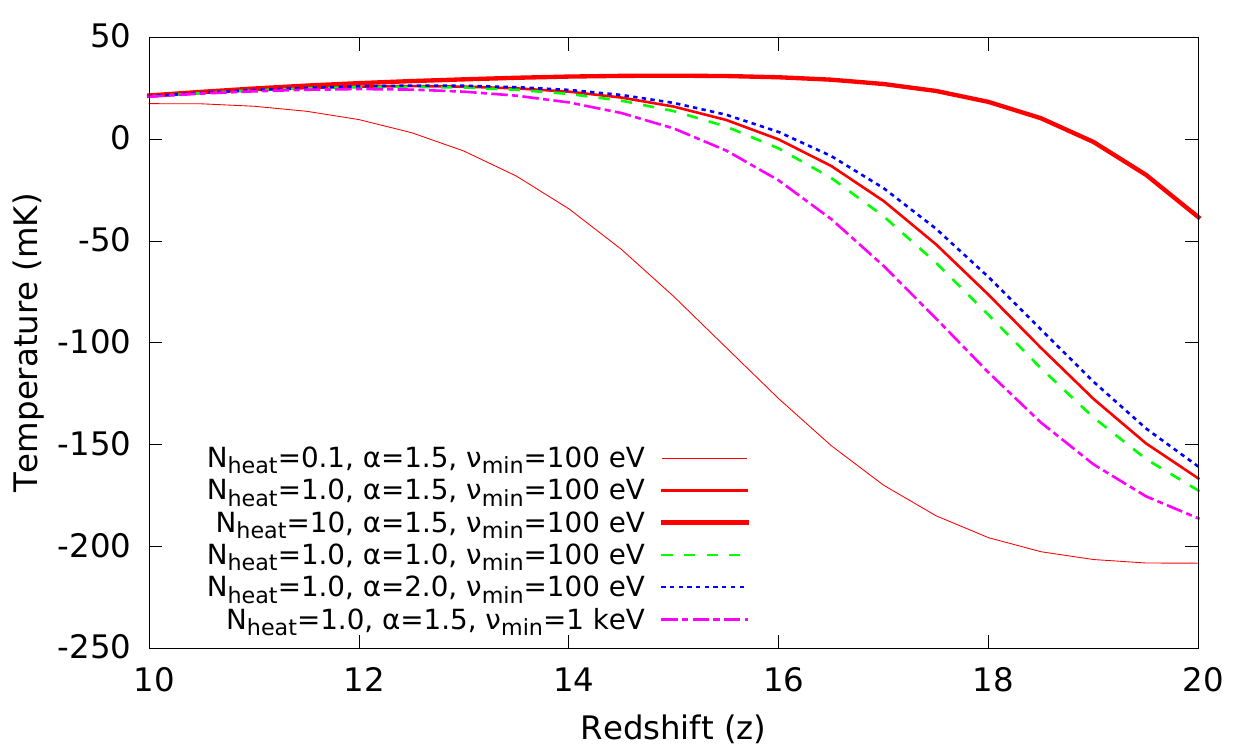}
		\caption{The evolution of global brightness temperature $\langle \Delta T_b \rangle$ (Eq.~(\ref{overallnorm})) is displayed for a range of  parameters for  $\zeta=10$.}
		\label{fig:glob_comp_10}
    \end{minipage}\hfill
\end{figure}


\begin{figure}
    \centering
    \begin{minipage}{0.49\textwidth}
		\centering
		\includegraphics[width=1.0\textwidth]{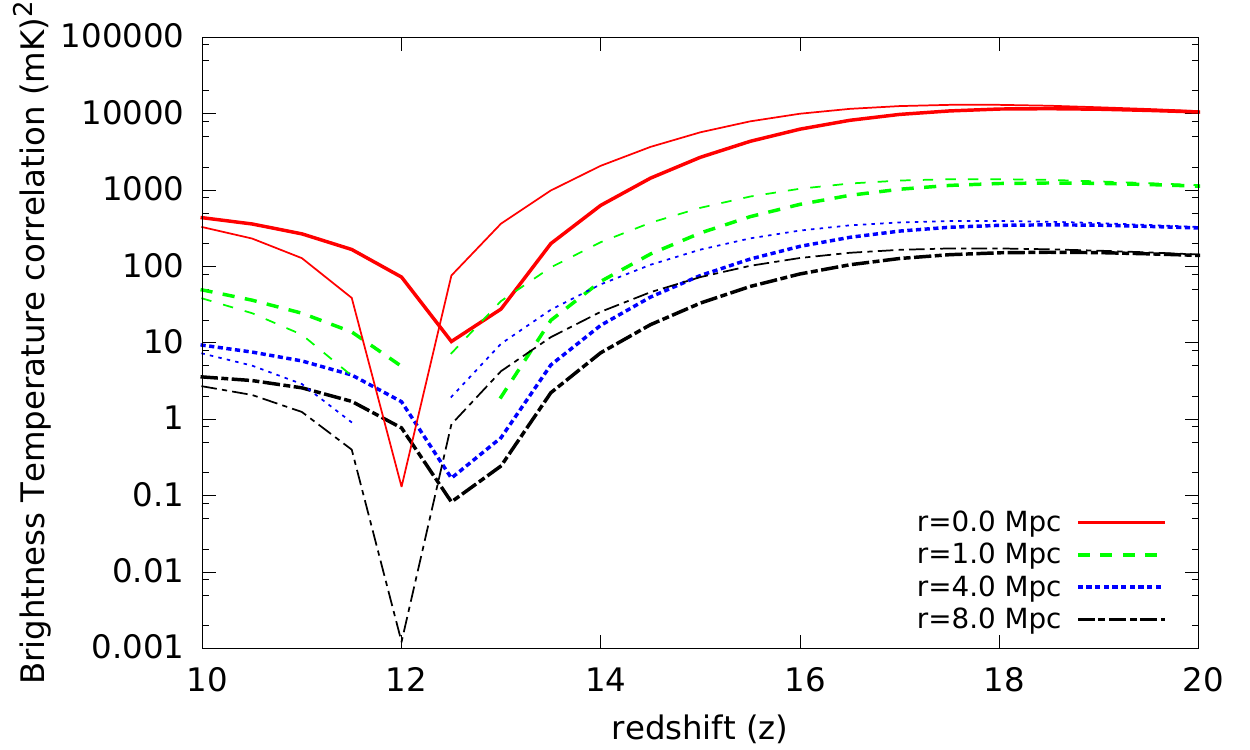}
    \end{minipage}\hfill
    \begin{minipage}{0.49\textwidth}
		\centering
		\includegraphics[width=1.0\textwidth]{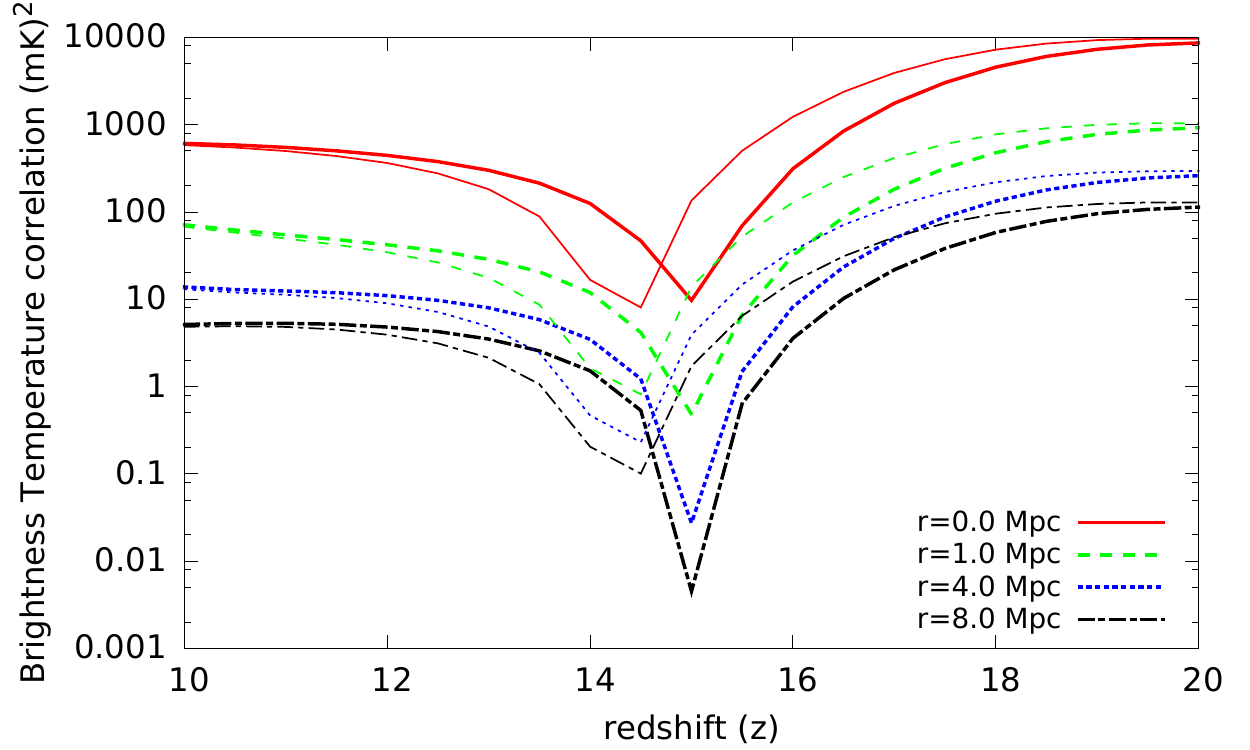}
    \end{minipage}\hfill
	\caption{Evolution of two-point  correlation function is  displayed for a range of scales (including the RMS corresponding to $r=0$) for $\alpha=1.5$,  $\zeta=10$,  and  two values of $\nu_{\rm min}$. The thick curves are for $\nu_{\rm min}= 100$ eV and the thin curves are for $\nu_{\rm min} = 1$ keV. The left  and right panels correspond to $N_{\text{heat}}=0.1$ and $N_{\text{heat}}=0.5$, respectively. }
	\label{fig:corr_evol_co}
    \begin{minipage}{0.49\textwidth}
		\centering
		\includegraphics[width=1.0\textwidth]{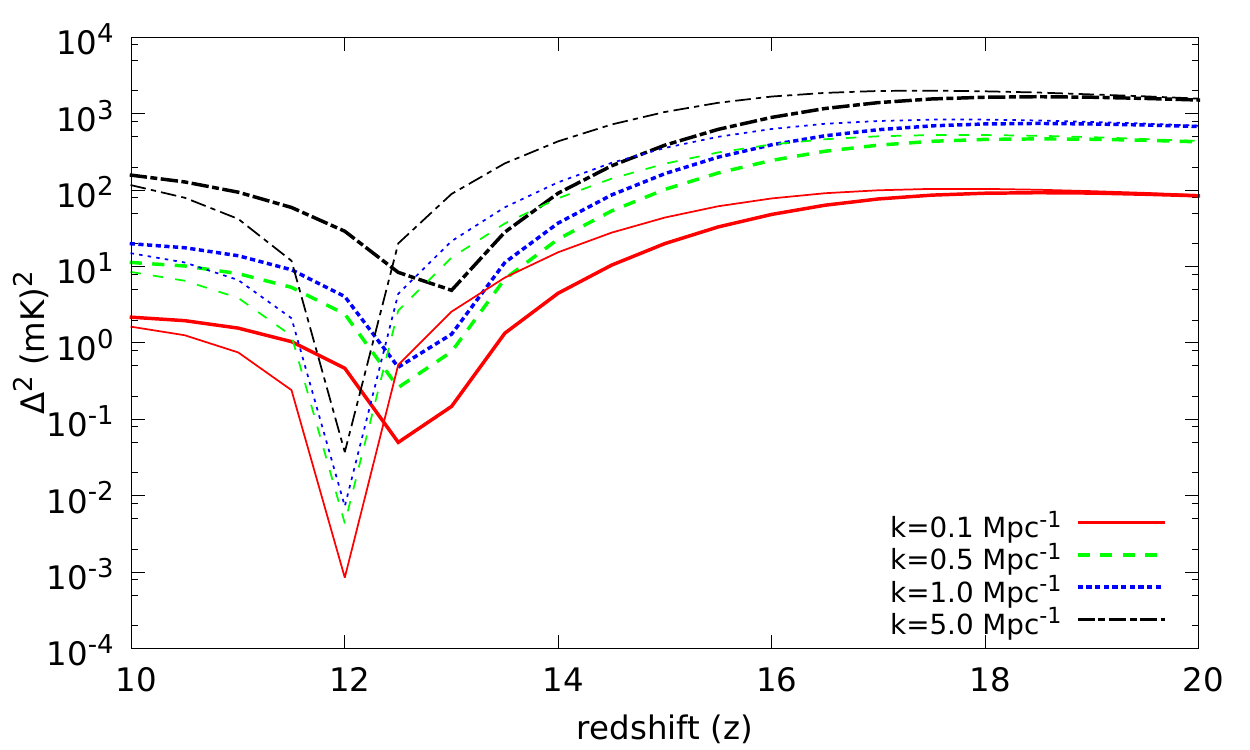}
	\end{minipage}\hfill
	\begin{minipage}{0.49\textwidth}
		\centering
		\includegraphics[width=1.0\textwidth]{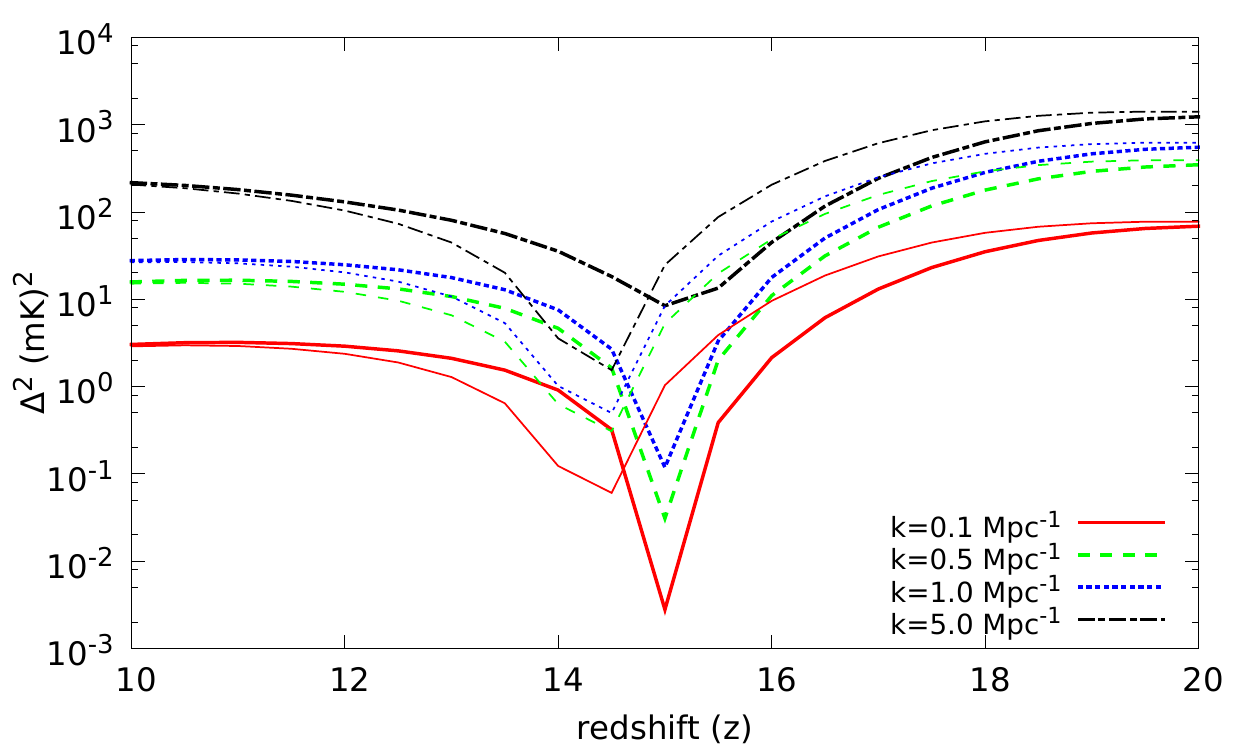}
	\end{minipage}\hfill
	\caption{Evolution of $\Delta^2 = k^3 P(k)/2\pi^2$ ($(mK)^2$) is  displayed for a range of scales for $\alpha=1.5$,  $\zeta=10$,  and  two values of $\nu_{\rm min}$. The thick curves are for $\nu_{\rm min}= 100$ eV and the thin curves are for $\nu_{\rm min} = 1$ keV. The left  and right panels correspond to $N_{\text{heat}}=0.1$ and $N_{\text{heat}}=0.5$, respectively. }
	\label{fig:PS_evol_co}

\end{figure}

In this sub-section, we present results for the complete model based on the 
$\Lambda$CDM model. This generalizes   the case discussed in the previous subsection in the following aspects: (a) there is a size distribution of ionization 
and heating bubbles (Figures~\ref{fig:ionbubbledis} and~\ref{fig:heat_dist_1.5_10_1}), (b) the heating bubbles have shallow profiles (Figure~\ref{fig:heat_prof_evol}), which makes
it harder to identify  the impact of sizes of heating bubbles on the  correlation function. 

{\it Dependence on modelling parameters and redshift}: In Figure~\ref{fig:corr_comp_z17}, we show  correlation functions for the complete model   for different choices of modelling parameters, using Eqs.~(\ref{overallnorm}),~(\ref{eq:massfun}),~(\ref{eq:fintemp}),~(\ref{eq:globtemp}),~(\ref{eq:corrfunexp}),~(\ref{eq:corrneu_case1}),~(\ref{eq:neus_cross1}), and~(\ref{eq:s1s2fin}). 
The correlation functions shown in the Figure are large  at  smalls scales (Eq.~(\ref{eq:corrss}))  and decrease as the distance between two points increases and approach Eq.~(\ref{eq:corrls}) at large scales. On intermediate scales the structure of correlation function is determined by the size distribution
of heated bubbles (e.g. Figure~\ref{fig:heat_dist_1.5_10_1}).

As the value of the spectral index  $\alpha$ is increased for a fixed $N_{\rm heat}$ there are more low frequency photons resulting in higher background temperature (Figure~\ref{fig:heat_prof_17_10_1_100}), this results in a decrease in the correlation function as it scales as $(1-s_b) = (1-\langle s \rangle)$ during this phase (Eq.~(\ref{eq:corrls})). For the same reason,  when $N_{\text{heat}}$ is increased, the correlation decreases.  Figures~\ref{fig:correvol_01} 
and~\ref{fig:correvol_10} show the correlation function    at different redshifts for different values of $N_{\rm heat}$. At high redshifts, the correlation function is large  owing to smaller background temperature. As the universe gets heated,  $s_b$ decreases reaching $f_n$  at a certain redshift
resulting in the vanishing of the correlation function at large scales (Eq.~(\ref{eq:corrls})). As 
the $s_b$  decreases further, the impact of partial heating 
disappears and   correlation function increases again owing to ionization 
inhomogeneities  and \ion{H}{1} signal is now observable in emission. We have 
already seen this transition for the single bubble, flat heating profile 
case (Figures~\ref{fig:flat_pro} and~\ref{fig:flat_pro_merge}).  In Figure~\ref{fig:glob_comp_10} we show the evolution of global \ion{H}{1} signal for 
a range of parameters. The dependence of the strength  
and the transition from absorption to emission  of the signal  on different parameters 
is as expected from the discussion in this section. 

We show the evolution of the correlation function for a range of scales in Figures~\ref{fig:corr_evol_co} for different values of $N_{\rm heat}$ and $\nu_{\rm min}$. For smaller value of $N_{\rm heat}$ and larger value of $\nu_{\rm min}$  the heating transition is delayed and the signal is larger during the era of partial heating, in line with the discussion
in the foregoing; this result is in qualitative agreement with similar analyses on delayed heating, e.g. \cite{LateHeat2}.

{\it Correlation scales}:  The scales at which we expect significant correlation
are determined by the distribution of sizes of heating bubbles, whose sizes
are determined by the sizes of ionization bubbles, $R_x$, the heating parameter
$N_{\rm heat}$, $\alpha$, and $\nu_{\rm min}$ (Eq.~(\ref{eq:fintemp})). In Figure~\ref{fig:heat_dist_1.5_10_1}, we show the size distribution of heating bubbles for a set of parameters. The heating bubble are larger than self-ionized bubbles by roughly a factor of 4.5 for this case and  at $z \simeq 14$ and the range of heating scales
lie in the range $2\hbox{--}7 \, \rm Mpc$. However, as noted above, the correlation function depends on the gradient of temperature and the fuzziness of the heating bubble doesn't allow one to readily identify these scales
in the correlation function. 

 Generically, a larger 
$N_{\rm heat}$ for a fixed $R_x$ results in larger heating bubbles and therefore 
causes correlation at larger scales. 
An increase in $\nu_{\rm min}$ causes 
shallow heating profiles, which results 
in reducing the gradient of temperature between the heating bubbles and the 
background, thereby reducing the correlation on a given scale for the same background temperature.  

In this 
paper,   we consider a set of models in which the parameters such as $N_{\rm heat}$ and $\nu_{\rm min}$ do not evolve. If these parameters are allowed to evolve the relation between
temperature inside heating bubbles and background temperature  would be more complicated. For instance, if $R_x$ is larger at an earlier epoch owing to 
the evolution of $\zeta$, the heating bubbles could be larger causing 
correlations on much larger scales than shown in this paper. 

{\it Can the merger  of heating bubbles introduce new correlation scales?}:
In this paper, we assume the centers of ionization bubbles to be uncorrelated.
It should be noted that the positions of ionizing sources is expected to be 
highly correlated and this effect has already been included in the definition
of self-ionized  bubbles. However, the mean bubble separation corresponds
to much larger scales at which the density correlation function for the 
$\Lambda$CDM model  is much 
smaller than unity,  and on such scales the  \ion{H}{1} the correlation 
function of the density  field is  expected to be smaller. The correlation
of  different ionizing 
centers is expected to follow the correlation function of the density field 
with a bias (e.g. \cite{2003moco.book.....D} and references therein). In this case 
the merging process is nearly  homogeneous and therefore does not introduce any 
new scales; its main effect, as noted above, is to wipe out the  correlation scales of heated bubbles.  It also follows that if the centers of self-ionized bubbles are assumed 
to be correlated, then, in principle, correlation at much larger scales can 
emerge. This can occur if the \ion{H}{1} field is highly biased with respect to the 
underlying density field (\cite{2015ApJ...802....8A}). We consider 
the case of correlated ionizing centers in Appendix~C and show that while  this 
effect doesn't alter our results qualitatively it  can introduce correlations
at new scales.

{\it power spectrum, comparison with existing results, and the topology of 
early reionization}: The early phase of reionization has been extensively 
studied in the literature using semi-analytic methods but primarily large-scale
simulations (e.g. \cite{2007MNRAS.376.1680P,2012Natur.487...70V,2013MNRAS.435.3001T,2013MNRAS.431..621M,LateHeat2,2015MNRAS.447.1806G,2014MNRAS.443..678P,2017MNRAS.464.3498F,2015PhRvL.114j1303F}). Most of these studies have been in the Fourier space. 
Our analysis suggests that real space correlation allows us to identify 
the physical processes more readily. Our results show that entire correlation
structure in real space can be written in terms of a single function: $C(.,.,.,.)$ and its limits 
given by the functions $E(.,.,.)$ and $D(.,.)$ (section~\ref{sec:Geometry}). 
Fourier transform with respect the first argument $r$ of this function yields
the power spectrum. In Figure~{\ref{fig:PS_evol_co}}, we show the evolution of the power spectrum for a range of Fourier modes $k$ for different values of $N_{\rm heat}$ and $\nu_{\rm min}$ (for details see Appendix~B6). 

Existing results show that, for $k \simeq 0.1\hbox{--}0.5 \, \rm Mpc^{-1}$, \footnote{Generally a wavenumber $k$  will contribute to a range of spatial scales; 
for making a comparison between real  space correlation function and power spectrum,
one can use the approximate conversion $r \simeq \pi/k$.}  during the era of early reionization where heating 
inhomogeneities dominate, there is a peak in the power spectrum which is followed by
a smaller peak at lower  redshifts when the inhomogeneities are dominated 
by ionization inhomogeneities (e.g. \cite{2007MNRAS.376.1680P,LateHeat2,2015MNRAS.447.1806G}). Our results (Figure~\ref{fig:PS_evol_co}) are in  agreement with this general picture. They 
are also in agreement  with analyses that have studied
the impact of partial heating  density perturbations at large scales (e.g. \cite{2013MNRAS.435.3001T,2013MNRAS.431..621M}) \footnote{At large scales, the correlation function approaches Eq.~(\ref{eq:corrls})
which is determined by density inhomogeneities;  its value could be 
enhanced by $(1-\langle s \rangle)^2$   at early times. Therefore, large scales
correlation function  at early times could be a reliable measure of density correlation
and  its statistical  anisotropy, in agreement with results
of \cite{2013MNRAS.435.3001T,2013MNRAS.431..621M}.} or the impact of late heating 
on the fluctuating component of the signal (\cite{LateHeat2}).  Many of  these analyses strongly suggest that 
the \ion{H}{1} signal is a robust probe of early X-ray heating, an inference 
our analysis adequately captures. We also  establish  the dependence of the signal on different parameters, 
which agrees with  existing results.  We do not attempt a more detailed comparison with the existing results  because it is hard to establish a one-to-one relation between parameters we use with  those  in the literature.

Our results are based on the assumption that the topology of the early 
reionization is given by Figure~\ref{fig:PH}: an ionized sphere surrounded by
a fuzzy heated region which merges smoothly into the background. The density 
perturbations determine the size of ionized regions, but the brightness 
temperature correlations on the scales of heating bubbles
are dominated by heating autocorrelation and heating-ionization cross-correlation. 

 The assumption of sphericity of ionized and heating regions for 
the computation of correlation function is reasonable 
even in the presence of density perturbations because the reionization process
is statistically isotropic and homogeneous\footnote{We neglect 
redshift-space distortion in the paper which renders the density field statistically anisotropic. However, so long as the ionization and heating sources
are isotropic their correlation is not affected by this anisotropy.}. However, the inclusion of density 
cross-correlation with other fields can alter the correspondence between 
the scale for a given set of physical parameter, e.g.  the scale of heating 
bubble is given roughly by the distance at which the optical depth of 
an X-ray photon reaches unity. We retain only background density for this 
computation but to be more exact we need to also include the impact of  density perturbations at this scale. Generally, the correlation of \ion{H}{1} density  field is small 
unless the \ion{H}{1} field is highly biased with respect to the underlying density 
field so this correction should be small at high redshift but will become  more important at smaller redshifts. Also we assume that the ionizing centers are 
uncorrelated, this assumption has greater validity at higher redshifts
when the mean separation between the centers is larger. At smaller redshift ($z\lesssim 12$, depending on the parameter $\zeta$), the excursion set formalism begins to break down since the ionization fraction becomes large and there is substantial overlap between ionization bubbles. In this regime, our results are not very accurate, however we still show results up to $z \simeq 10$, to emphasize the transition from the era of domination of heating inhomogeneities to ionization inhomogeneities.

While N-body simulations assume paramount importance if the \ion{H}{1} density 
field at large redshifts is to be directly imaged, all  of the ongoing experiments that seek to detect this signal rely upon statistical detection of this signal. Our method cannot  predict the shape of individual regions but allows 
us to compute  the statistics of the \ion{H}{1} field.  Another advantage with analytic
estimates is that they  are computationally inexpensive as compared to simulations. 
Given the uncertainty in the early heating phase of the universe, our analysis can be 
used to compare the observed signal for a multiple set of parameters 
and better understand their degeneracies at a fraction of computational cost
needed to carry out an N-body simulation.  For instance, the \ion{H}{1} signal in early heating regime depends on the gradient of temperature in and across 
heated regions whose sizes  depend on  multiple physical processes.  It would be of great interest 
to determine whether the future data can distinguish between these  different physical  processes.

{\it Detectability of the signal}: Many  operational  (e.g. LOFAR,  MWA, PAPER, GMRT)  and upcoming radio interferometers (HERA, SKA)  have the capability to detect the 
fluctuating component of the \ion{H}{1} signal in the redshift range $8 <z < 25$ (for details e.g. \cite{2015aska.confE...3A,2015aska.confE...1K,2014MNRAS.439.3262M}). It is customary in the literature to present  the  sensitivity of radio interferometers for the detection of \ion{H}{1} signal 
 in terms of power spectrum, partly  because the radio interferometers measure the 
Fourier component of the \ion{H}{1} signal.  However, these estimates can be extended to 
image plane  (which is often a byproduct of the analysis pipeline e.g. \cite{2017ApJ...838...65P} for LOFAR) or real-space correlation functions used for computation of the signal in this paper (e.g. \cite{2008ApJ...673....1S}).   We discuss here the expected 
sensitivity of SKA1-LOW (\cite{2015aska.confE...1K}). For a deep survey with  SKA1-LOW, the error
on the power spectrum ($\Delta^2 \equiv k^3 P(k)/(2\pi^2)$) is expected to vary
from $0.1 \, \rm (mK)^2$ at $z \simeq 9$ to $5 \, \rm (mK)^2$ at 
$z \simeq 25$ for $k = 0.1 \, \rm Mpc^{-1}$. At $z \simeq 16$, the expected 
error is $2 \, \rm (mK)^2$ increasing to $10 \, \rm (mK)^2$ for $k \simeq 0.5 \, \rm Mpc^{-1}$ (for details see Figure~2 of  \cite{2015aska.confE...1K}).  Direct comparison with Figure~\ref{fig:PS_evol_co} shows that the \ion{H}{1} signal at $z\simeq 16$ can be detected with a signal-to-noise varying  from 50 to 10  for $0.1 < k < 0.5 \, \rm Mpc^{-1}$. 

We can get similar estimates from the signal in real space by using $r \simeq \pi/k$.   \footnote{Angular scale above which the \ion{H}{1} signal can be reliably measured for most  ongoing and upcoming radio interferometers is  a few arcminutes;  $1'$ corresponds to nearly 
3~Mpc (comoving) at $z \simeq 15$ or  these telescopes are sensitive to linear 
scales larger than  $5\hbox{--}10 \, \rm  Mpc$ (comoving). However, these telescopes 
have frequency resolution which correspond to  much smaller linear scales, e.g. 
MWA's frequency resolution of 40~kHZ corresponds to nearly 1~Mpc (comoving) along the  line of sight. Or 
the 3-d \ion{H}{1} signal is probed with different resolution on the sky plane as 
compared to the line of sight.}

\section{Summary and conclusions} \label{sec:sumcon}

The main aim of this paper is to present a new analytic formalism to 
study the phase of EoR that is dominated by partial heating. 

The main ingredients of our analytic model are: (a) correlation of the \ion{H}{1} 
density field is given by the $\Lambda$CDM model. At large scales, this 
correlation dominates (Eq.~(\ref{eq:corrls})). (b) correlation of  ionization is 
determined by the size distribution of self-ionized bubbles. The definition of self-ionized bubbles takes into account the clustering of halos as they  form in high density regions. The cross-correlation between density and ionization inhomogeneities is neglected in our work. Both (a) and (b) have been extensively studied 
both analytically and numerically in the literature.   (c)  modelling of 
heating inhomogeneities using near- and far-zone around the centers
of  self-ionized regions. While the phase of partial heating has 
been studied in the literature, this formulation is new and allows
us to compute the statistical quantities related to the \ion{H}{1} signal, (d) computation of two-point correlation functions in real space for a sharp ionized 
region surrounded by a fuzzy heated bubble. We develop a formalism to compute
these functions. In particular, we take into account the heating autocorrelation and heating-ionization cross-correlations while neglecting the density-heating cross-correlation. We also take into consideration both the correlation 
for a single bubble and for multi-bubbles, assuming the centers of self-ionized regions to be uncorrelated. We also explicitly show that our formalism reduces
to the correct form in different limits  (discussion following Eq.~(\ref{eq:flatpro_fin})). 
In many  extensions of $\Lambda$CDM model the power at small scales can 
differ substantially from the usual model (e.g.  \cite{2009JCAP...11..021S,2016JCAP...04..012S}), our formalism can be extended to such models by generating 
the size distribution of self-ionized bubbles using the matter power spectra of these models.

We model the ionization and heating using four parameters: $\zeta$, which 
determines the ionization history of EoR and is constrained by Planck and WMAP
results and three parameters to model heating---$N_{\rm heat}$, the number of X-ray photons per stellar baryons, $\alpha$, the spectral index of X-ray photons, 
and $\nu_{\rm min}$, the lowest frequency of X-ray photons. We study the 
impact of these parameters on the correlation functions and find reasonable 
agreement with existing results. 

In this paper, we assume homogeneous coupling between  Lyman-$\alpha$ photons
and neutral hydrogen through Wouthuysen-Field effect such that $T_\alpha = T_K$. As discussed in section~\ref{sec:lyalpha}, this is a good assumption for $z < 20$ but we expect 
imperfect coupling at higher redshifts which could create inhomogeneities 
in the \ion{H}{1} signal, resulting in another peak in the evolution of the signal at multiple scales  (e.g.  \cite{2008ApJ...684...18C,2007MNRAS.376.1680P,2015aska.confE...3A}). These inhomogeneities arise owing to the absorption 
 of photons between 
Lyman-$\beta$ and Lyman-limit closer to the ionizing sources,  unlike photons 
between Lyman-$\alpha$ and Lyman-$\beta$ discussed in section~\ref{sec:lyalpha}, and these inhomogeneities can be studied using formalism developed in this 
paper. We hope to return to this issue in the near future.

Given the uncertainty in the heating history during EoR and its impact on the 
\ion{H}{1} signal, our analytic formulation  allows us to  isolate the impact of 
different physical parameters and also underline their degeneracies. Future data in the redshift range of interest, $10 < z< 20$, is likely to put strong 
constraints on the physical processes during this era. Our work  is one 
step in the direction of understanding the data better. 
\acknowledgments
\section*{Acknowledgment}
We would like to thank Saurabh Singh, Steven Furlanetto and Jordan Mirocha for useful discussions and comments on the manuscript.
\pagebreak

\appendix
\section{Probability}
	\begin{align}
		P(A|B)&=\frac{P(A\cap B)}{P(B)} \label{eq:B2} \\
		P((A\cap B)|C) &=P(A|(B\cap C))\; P(B|C) \label{eq:B3} \\
		P(A\cap B) &=P(A)-P(A\cap \tilde B) \label{eq:AandB} 
	\end{align}
\section{Geometry} \label{sec:Geometry}
\subsection{$A(R_1,R_2,d)$}
Given two spheres of radius $R_1$ and $R_2$, the surface area of the sphere of radius $R_1$ that lies inside sphere of radius $R_2$ is 
\footnote{mathworld.wolfram.com/Sphere-SphereIntersection.html \\
	http://mathworld.wolfram.com/Zone.html}:
	\begin{equation}
		A(R_1,R_2,d)=\left\{
		\begin{array}{cl}
			0 &\quad d>R_1+R_2\\
			0 &\quad d<R_1-R_2\\
			4\pi R_1^2&\quad d<R_2-R_1\\
			\frac{\pi R_1 (R_2-R_1+d)(R_2+R_1-d)}{d} &\quad \text{Otherwise}
		\end{array} \right. \label{eq:AxRd}
	\end{equation}
where $d=$ distance between two sphere centers.

\subsection{$V(R_1,R_2,d)$}
Given two spheres of radius $R_1$ and $R_2$, the overlapped volume is 
\begin{equation}
V(R_1,R_2,d)=\left\{
\begin{array}{cl}
0 &\quad d>R_1+R_2\\
\frac{4\pi}{3}R_2^2 &\quad d<R_1-R_2\\
\frac{4\pi}{3}R_1^2 &\quad d<R_2-R_1\\
\frac{\pi (R_2+R_1-d)^2 (d^2+2d(R_1+R_2)-3(R_1-R_2)^2)}{12d} &\quad \text{Otherwise}
\end{array} \right. \label{eq:VxRd}
\end{equation}
where $d=$ distance between two sphere centers.
	
\subsection{$C(x,P,Q,R)$} \label{sec:defc}
If point 1 is located between distance $P$ and $Q$ $(P<Q)$ from the center of a sphere, then $C(x,P,Q,R)$ is the probability that its neighbour point 2 at distance $x$ from point 1 is located outside the concentric sphere of radius $R$. We assume $a=$ distance of point 1 from the center of the sphere.

\textbf{(a)} $x \leq R-Q$: None of the neighbours of point 1 are outside the sphere of radius $R$, giving $ C(x,P,Q,R) = 0$.

\textbf{(b)} $x \leq P-R$: All the neighbours of point 1 are outside the sphere of radius $R$, giving $ C(x,P,Q,R) = 1$.

\textbf{(c)} $x \geq R+Q$: All the neighbours of point 1 are outside the sphere of radius $R$, giving $ C(x,P,Q,R) = 1 $.

\textbf{(d)} $R-Q \leq x \leq R-P$: Taking $S=R-x$, we can see that all the neighbours of the points between distance $P$ and $S$ are inside the sphere of radius $R$. Therefore we have,
	\begin{align}
		C(x,P,Q,R)=\left[\frac{\frac{4\pi}{3}(S^3-P^3)}{\frac{4\pi}{3}(Q^3-P^3)}\right](0)+ \int_{S}^{Q}\left(1-\frac{A(x,R,a)}{A_x}\right) \frac{4\pi a^2}{\frac{4\pi}{3}(Q^3-P^3)}\mathrm{d}a \nonumber
	\end{align}
Where, $A_x = 4\pi x^2$ is the area of a sphere of radius $x$.

	\textbf{(d.a)} If $x \leq Q-R$: Taking $T = x+R$, we see that all neighbours of points between distance $T$ and $Q$ are outside the sphere of radius $R$.
	\begin{align}
		C(x,P,Q,R)&=\int_{S}^{T}\left(1-\frac{A(x,R,a)}{A_x}\right) \frac{4\pi a^2}{\frac{4\pi}{3}(Q^3-P^3)}\mathrm{d}a + \int_{T}^{Q}\left(1-\frac{0}{A_x}\right) \frac{4\pi a^2}{\frac{4\pi}{3}(Q^3-P^3)}\mathrm{d}a \nonumber \\
			&= \frac{Q^3-R^3}{Q^3-P^3} \nonumber
	\end{align}

	\textbf{(d.b)} Otherwise,
	\begin{align}
		C(x,P,Q,R) &= \frac{1}{2} \frac{Q^3-(R-x)^3}{Q^3-P^3} + \frac{3}{8x} \frac{Q^2-(R-x)^2}{Q^3-P^3} \left[\frac{Q^2+(R-x)^2}{2}+(x^2-R^2)\right] \nonumber
	\end{align}
	
\textbf{(e)} $P-R \leq x \leq Q-R$: Taking $S=R+x$, we can see that all the neighbours of the points between distance $S$ and $Q$ are outside the sphere of radius $R$. Therefore we have,
	\begin{align}
		C(x,P,Q,R)=\left[\frac{\frac{4\pi}{3}(Q^3-S^3)}{\frac{4\pi}{3}(Q^3-P^3)}\right](1)+ \int_{P}^{S}\left(1-\frac{A(x,R,a)}{A_x}\right) \frac{4\pi a^2}{\frac{4\pi}{3}(Q^3-P^3)}\mathrm{d}a \nonumber
	\end{align}

	\textbf{(e.a)} If $x \geq P+R$: $T = x-R$, we can see that all the neighbours of the points between distance $P$ and $Q$ are outside the sphere of radius $R$. Therefore we have,
	\begin{align}
		C(x,P,Q,R)&=\left[\frac{\frac{4\pi}{3}(Q^3-S^3)}{\frac{4\pi}{3}(Q^3-P^3)}\right](1)+ \int_{P}^{T}\left(1-\frac{0}{A_x}\right) \frac{4\pi a^2}{\frac{4\pi}{3}(Q^3-P^3)}\mathrm{d}a +\int_{T}^{S}\left(1-\frac{A(x,R,a)}{A_x}\right) \frac{4\pi a^2}{\frac{4\pi}{3}(Q^3-P^3)}\mathrm{d}a \nonumber \\
			&=	1 - \frac{R^3}{Q^3-P^3} \nonumber
	\end{align}
	
	\textbf{(e.b)} For $x \leq P+R$, there are two probabilities. If $x<R-P$, we are left with the case \textbf{(d.b)}. Otherwise,
	\begin{align}
		C(x,P,Q,R) &= \frac{Q^3-(R+x)^3/2-P^3/2}{Q^3-P^3} + \frac{3}{8x} \frac{(R+x)^2-P^2}{Q^3-P^3} \left[\frac{(R+x)^2+P^2}{2}+(x^2-R^2)\right] \nonumber
	\end{align}

\textbf{(f)} $R+P \leq x \leq R+Q$: Taking $S=x-R$, we can see that all the neighbours of the points between distance $S$ and $P$ are outside the sphere of radius $R$. Therefore we have,
	\begin{align}
		C(x,P,Q,R)=\left[\frac{\frac{4\pi}{3}(S^3-P^3)}{\frac{4\pi}{3}(Q^3-P^3)}\right](1)+ \int_{S}^{Q}\left(1-\frac{A(x,R,a)}{A_x}\right) \frac{4\pi a^2}{\frac{4\pi}{3}(Q^3-P^3)}\mathrm{d}a \nonumber
	\end{align}
	
	\textbf{(f.a)} If $x \leq Q-R$: $T=x+R$, we can see that all the neighbours of the points between distance $T$ and $Q$ are outside the sphere of radius $R$. This is the same case as \textbf{(e.a)} (with $T$ and $S$ exchanged). Therefore we have,
	\begin{align}
		C(x,P,Q,R) &=	1 - \frac{R^3}{Q^3-P^3} \nonumber
	\end{align}
	
	\textbf{(f.b)} Otherwise,
	\begin{align}
		C(x,P,Q,R) &= \frac{(x-R)^3-P^3}{Q^3-P^3} + \frac{1}{2} \frac{Q^3-(x-R)^3}{Q^3-P^3} + \frac{3}{8x} \frac{Q^2-(x-R)^2}{Q^3-P^3} \left[\frac{Q^2+(x-R)^2}{2}+(x^2-R^2)\right] \nonumber
	\end{align}
	
\textbf{(g)} For the last case, when $x \geq Q-R$, $x \geq R-P$, $x \leq P+R$,
	\begin{align}
		C(x,P,Q,R)&= \int_{P}^{Q} \left(1-\frac{A(x,R,a)}{A_x}\right) \frac{4\pi a^2}{\frac{4\pi}{3}(Q^3-P^3)} \mathrm{d}a \nonumber \\
			&=	\frac{1}{2}+\frac{3}{8x}\frac{P+Q}{P^2+PQ+Q^2}\left[\frac{P^2+Q^2}{2}+(x^2-R^2)\right] \nonumber
	\end{align}
	
	\begin{equation}
		C(x,P,Q,R)=\left\{
		\begin{array}{cl}
			0 &\quad x \leq R-Q\\
			1 &\quad x \leq P-R\\
			1 &\quad x \geq R+Q\\
			\frac{1}{2} \frac{Q^3-(R-x)^3}{Q^3-P^3} + \frac{3}{8x} \frac{Q^2-(R-x)^2}{Q^3-P^3} \left[\frac{Q^2+(R-x)^2}{2}+(x^2-R^2)\right] & R-Q \leq x \leq R-P,\; x > Q-R\\
			\frac{Q^3-R^3}{Q^3-P^3} &R-Q \leq x \leq R-P,\; x \leq Q-R\\
			\frac{1}{2} \frac{2Q^3-(R+x)^3-P^3}{Q^3-P^3} + \frac{3}{8x} \frac{(R+x)^2-P^2}{Q^3-P^3} \left[\frac{(R+x)^2+P^2}{2}+(x^2-R^2)\right] & |P-R| \leq x \leq Q-R,\;x < P+R\\
			1 - \frac{R^3}{Q^3-P^3} & P+R \leq x \leq Q-R,\;\\
			\frac{1}{2} \frac{(x-R)^3+Q^3-2P^3}{Q^3-P^3} + \frac{3}{8x} \frac{Q^2-(x-R)^2}{Q^3-P^3} \left[\frac{Q^2+(x-R)^2}{2}+(x^2-R^2)\right] & R+P \leq x \leq R+Q,\;x > Q-R\\
			\frac{1}{2} + \frac{3}{8x} \frac{P+Q}{P^2+PQ+Q^2} \left[\frac{P^2+Q^2}{2}+(x^2-R^2)\right] & R-P \leq x \leq P+R,\; x \geq Q-R
		\end{array} \right.
	\end{equation}

\subsection{$E(x,Q,R)$}
	If point 1 is inside a sphere of radius $Q$, then the probability that its neighbour (point 2) at distance $x$ is outside the concentric sphere of radius $R$ is $E(x, Q, R)$. We assume $a=$ distance of point 1 from the center of the sphere. 
	
	\textbf{(a)} $R>Q$ and $x<R-Q$: All neighbors of point 1 are inside the sphere of radius $R$, giving $ E(x,Q,R) = 0$.
	
	\textbf{(b)} $R>Q$ and $R-Q<x<R$: $R-x<Q$. If $a <R-x$, all neighbors of point 1 are inside the sphere of radius $R$. For $a>R-x$, we can use \eqref{eq:AxRd}.
		\begin{align}
			E(x,Q,R) &= \left[\frac{\frac{4\pi}{3}(R-x)^3}{\frac{4\pi}{3}Q^3}\right](0) + \int_{R-x}^{Q} \left(1-\frac{A(x,R,a)}{A_x}\right) \frac{4\pi a^2}{\frac{4\pi}{3}Q^3} \mathrm{d}a \nonumber \\
				&= \frac{1}{2} - \frac{R^3}{2Q^3} + \frac{3R^4}{16Q^3x} - \frac{3R^2}{8Qx} + \frac{3Q}{16x} + \frac{3R^2x}{8Q^3} + \frac{3x}{8Q} - \frac{x^3}{16Q^3} \nonumber
		\end{align}
		
	\textbf{(c)} $R>Q$ and $R<x<R+Q$: $x-R<Q$. If $a<x-R$, all neighbors of point 1 are outside the sphere of radius $R$.
		\begin{align}
			E(x,Q,R) &= \left[\frac{\frac{4\pi}{3}(x-R)^3}{\frac{4\pi}{3}Q^3}\right](1) + \int_{x-R}^{Q} \left(1-\frac{A(x,R,a)}{A_x}\right) \frac{4\pi a^2}{\frac{4\pi}{3}Q^3} \mathrm{d}a \nonumber \\
				&= \frac{1}{2} - \frac{R^3}{2Q^3} + \frac{3R^4}{16Q^3x} - \frac{3R^2}{8Qx} + \frac{3Q}{16x} + \frac{3R^2x}{8Q^3} + \frac{3x}{8Q} - \frac{x^3}{16Q^3} \nonumber
		\end{align}
		
	\textbf{(d)} $R+Q<x$: All neighbors of point 1 are outside the sphere of radius $R$, giving $
			E(x,Q,R) =1 $.
			
	\textbf{(e)} $Q>R$, $x<Q-R$ and $x<R$: $R-x>0$ and $x+R<Q$. If $a<R-x$, all neighbors of point 1 are inside the sphere of radius $R$. If $a>R+x$, all neighbors of point 1 are inside the sphere of radius $R$.
		\begin{align}
			E(x,Q,R) &= \left[\frac{\frac{4\pi}{3}(R-x)^3}{\frac{4\pi}{3}Q^3}\right](0) + \int_{R-x}^{x+R} \left(1-\frac{A(x,R,a)}{A_x}\right) \frac{4\pi a^2}{\frac{4\pi}{3}Q^3} \mathrm{d}a +\left[\frac{\frac{4\pi}{3}(R+x)^3}{\frac{4\pi}{3}Q^3}\right](1) \nonumber \\
				&= 1 - \frac{R^3}{Q^3} \nonumber
		\end{align}
		
	\textbf{(f)} $Q>R$, $x<Q-R$ and $R<x<Q+R$: $x-R>0$ and $x+R<Q$. If $a<x-R$, all neighbors of point 1 are outside the sphere of radius $R$. If $a>R+x$, all neighbors of point 1 are inside the sphere of radius $R$.
		\begin{align}
			E(x,Q,R) &= \left[\frac{\frac{4\pi}{3}(x-R)^3}{\frac{4\pi}{3}Q^3}\right](1) + \int_{x-R}^{x+R} \left(1-\frac{A(x,R,a)}{A_x}\right) \frac{4\pi a^2}{\frac{4\pi}{3}Q^3} \mathrm{d}a +\left[\frac{\frac{4\pi}{3}(R+x)^3}{\frac{4\pi}{3}Q^3}\right](1) \nonumber \\
				&= 1 - \frac{R^3}{Q^3} \nonumber
		\end{align}
		
	\textbf{(g)} $Q>R$, $x>Q-R$ and $x<R$: $R-x>0$. If $a<R-x$, all neighbors of point 1 are inside the sphere of radius $R$.
		\begin{align}
			E(x,Q,R) &= \left[\frac{\frac{4\pi}{3}(R-x)^3}{\frac{4\pi}{3}Q^3}\right](0) + \int_{R-x}^{Q} \left(1-\frac{A(x,R,a)}{A_x}\right) \frac{4\pi a^2}{\frac{4\pi}{3}Q^3} \mathrm{d}a \nonumber \\
				&= \frac{1}{2} - \frac{R^3}{2Q^3} + \frac{3R^4}{16Q^3x} - \frac{3R^2}{8Qx} + \frac{3Q}{16x} + \frac{3R^2x}{8Q^3} + \frac{3x}{8Q} - \frac{x^3}{16Q^3} \nonumber
		\end{align}
		
	\textbf{(h)} $Q>R$, $x>Q-R$ and $R<x<Q+R$: $x-R>0$. If $a<x-R$, all neighbors of point 1 are outside the sphere of radius $R$.
		\begin{align}
			E(x,Q,R) &= \left[\frac{\frac{4\pi}{3}(x-R)^3}{\frac{4\pi}{3}Q^3}\right](1) + \int_{x-R}^{Q} \left(1-\frac{A(x,R,a)}{A_x}\right) \frac{4\pi a^2}{\frac{4\pi}{3}Q^3} \mathrm{d}a \nonumber \\
				&= \frac{1}{2} - \frac{R^3}{2Q^3} + \frac{3R^4}{16Q^3x} - \frac{3R^2}{8Qx} + \frac{3Q}{16x} + \frac{3R^2x}{8Q^3} + \frac{3x}{8Q} - \frac{x^3}{16Q^3} \nonumber
		\end{align}
		
	\begin{equation}
		E(x,Q,R)=\left\{
		\begin{array}{cl}
			0 & \quad x<R-Q \\
			1 - \frac{R^3}{Q^3} & \quad R-Q<x<Q-R \\
			\frac{1}{2} - \frac{R^3}{2Q^3} + \frac{3R^4}{16Q^3x} - \frac{3R^2}{8Qx} + \frac{3Q}{16x} + \frac{3R^2x}{8Q^3} + \frac{3x}{8Q} - \frac{x^3}{16Q^3} & \quad |R-Q|<x<R+Q \\
			1 &\quad x>R+Q
		\end{array} \right.	
	\end{equation}
	This can also be derived using $E(x,Q,R)=C(x,0,Q,R)$.

\subsection{$D(x, R)$}
	If point 1 is inside a sphere of radius $R$, then the probability that its neighbour (point 2) at distance $x$ is outside the sphere is $D(x,R)$. We assume $a=$ distance of point 1 from the center of the sphere. 
	
	\textbf{(a)} $x<R$: If  $a<R-x$, then all the neighbors of point 1 are inside the sphere. For $a>R-x$, we can use \eqref{eq:AxRd}.
		\begin{align}
			D(x, R)&=\left[\frac{\frac{4\pi}{3}(R-x)^3}{\frac{4\pi}{3}R^3}\right](0)+ \int_{R-x}^{R}\left(1-\frac{A(x,R,a)}{A_x}\right) \frac{4\pi a^2}{\frac{4\pi}{3}R^3}\mathrm{d}a \nonumber \\
				&= 	\frac{3x}{4R}-\frac{x^3}{16R^3}\nonumber
		\end{align}
		
	\textbf{(b)} $R<x<2R$: If $a<x-R$, then all neighbors of point 1 are outside the sphere. For $a>x-R$, we can use \eqref{eq:AxRd}.
		\begin{align}
			D(x,R)&=\left[\frac{\frac{4\pi}{3}(x-R)^3}{\frac{4\pi}{3}R^3}\right](1)+ \int_{x-R}^{R}\left(1-\frac{A(x,R,a)}{A_x}\right) \frac{4\pi a^2}{\frac{4\pi}{3}R_x^3}\mathrm{d}a \nonumber \\
				&=	\frac{3x}{4R}-\frac{x^3}{16R^3} \nonumber
		\end{align}
		
	\textbf{(c)} $2R<x$: All neighbors of point 1 are outside the sphere, giving $ D(x,R)=1  $.
	
	\begin{equation}
		D(x,R)=\left\{
		\begin{array}{cl}
			\frac{3x}{4R}-\frac{x^3}{16R^3} &\quad x<2R\\
			1 &\quad x>2R
		\end{array} \right.
	\end{equation}
	This can also be derived using $D(x,R)=E(x,R,R)=C(x,0,R,R)$.
	
\subsection{Fourier Transform of $1-C(x,P,Q,R)$}
	As $C(x,P,Q,R)$  goes to unity  at large $x$,   we re-arrange the terms of correlation function so that we only have to deal with $1-C(x,P,Q,R)$, which vanishes in this limit.
	
	To take Fourier transform of $1-C(x,P,Q,R)$, we can take six sub-cases: (1) $R \leq P<Q$ and $P+R \leq Q-R$; (2) $R \leq P <Q$ and $ Q-R \leq P+R$; (3) $P \leq R \leq Q$ and $Q-R \leq R-P$; (4) $P \leq R \leq Q$ and $R-P \leq Q-R \leq R+P$; (5) $P \leq R \leq Q$ and $R+P \leq Q-R $; (6) $P < Q \leq R$. We can take cases of $C(x,P,Q,R)$ for different limits for different cases. However, simplifying them, we get the following common form: 
	\begin{align}
	{\rm F.T.}(1-C(x,P,Q,R)) &= \frac{3}{64 k^4 \pi^4 (P^3 - Q^3)} \nonumber \\
	&\quad (- (1 + 2k^2 \pi^2 (Q + R)^2) {\rm Cos}(2 k \pi(Q - R)) \nonumber \\
	&\quad\; - (1 + 2k^2 \pi^2 (P - R)^2) {\rm Cos}(2 k\pi(P + R)) \nonumber \\
	&\quad\; + (1 + 2k^2 \pi^2 (Q - R)^2) {\rm Cos}(2 k\pi(Q + R)) \nonumber \\
	&\quad\; + (1 + 2k^2 \pi^2 (P + R)^2) {\rm Cos}(2 k\pi(P - R)) \nonumber \\
	&\quad + 2 k \pi ( + (-1 + 2k^2 \pi^2 (P-R)^2)(R+P) {\rm Sin}(2 k\pi(P+R)) \nonumber \\
	&\qquad\qquad\;- (-1 + 2k^2 \pi^2 (Q-R)^2)(Q+R) {\rm Sin}(2 k\pi(Q+R)) \nonumber \\
	&\qquad\qquad\;- (-1 + 2k^2 \pi^2 (P+R)^2)(P-R) {\rm Sin}(2 k\pi(P-R)) \nonumber \\
	&\qquad\qquad\;+ (-1 + 2k^2 \pi^2 (Q+R)^2)(Q-R) {\rm Sin}(2 k\pi(Q-R))) \nonumber \\
	&\quad + 8k^4 \pi^4 ( -(P^2-R^2)^2 ( {\rm Ci}(P+R) - {\rm Ci}(|P-R|)) \nonumber \\
	&\qquad\qquad\quad +(Q^2-R^2)^2 ( {\rm Ci}(Q+R) - {\rm Ci}(|Q-R|))) )\nonumber 
	\end{align}
	The different cases mentioned above  give different limits for Cos Integral (Ci) function. They can be rearranged  to make  sure that the argument of Ci function is always positive. To ensure the power spectrum is a positive definite function, we extract the symmetric part of the resulting expressions.  
	
\section{Correlation of Bubble Centers} \label{sec:Correlations}
\subsection{Correlation of Ionized and Heated Regions} \label{sec:corr_ion}
	If there are ionization bubbles of only radius $R_x$ and if we assume there is no correlation between ionization bubble centers, then the probability of a point at distance $r$ from the center of an ionized bubble to be ionized is,
		\begin{align}
			P_0 &=f_i=\frac{4\pi}{3}R_x^3 N(R_x) \nonumber
		\end{align}
	Where $N(R_x)=$ Number density of bubbles. Now, if we assume that there is correlation between ionization bubble centers, then this probability changes. The probability of a point located at distance $r$ in a shell of width $\Delta r$ to be ionized is, 
		\begin{align}
			P &= \int_{r-\Delta r/2-R_x}^{r+\Delta r/2+R_x} \frac{4\pi x^2 (1+F(x)) N(R_x) \mathrm{d}x}{\frac{4\pi}{3} \left[\left(r+\frac{\Delta r}{2}\right)^3- \left(r-\frac{\Delta r}{2}\right)^3\right]} \left[V\left(R_x, r+\frac{\Delta r}{2}, x \right) - V\left(R_x, r-\frac{\Delta r}{2}, x \right)\right] \nonumber
		\end{align}
	where $F(x)$ is the correlation function of ionization bubble centers at distance $x$. This gives, 
		\begin{align}
			P &= \frac{4\pi}{3}N(R_x)R_x^3 + \int_{r-\Delta r/2-R_x}^{r+\Delta r/2+R_x} \frac{4\pi x^2 F(x) N(R_x) \mathrm{d}x}{\frac{4\pi}{3} \left[\left(r+\frac{\Delta r}{2}\right)^3- \left(r-\frac{\Delta r}{2}\right)^3\right]} \left[V\left(R_x, r+\frac{\Delta r}{2}, x \right) - V\left(R_x, r-\frac{\Delta r}{2}, x \right)\right] \nonumber
		\end{align}
	For $\Delta r \to 0$, we have,
		\begin{align}
				P &= f_i\left(1 + \int_{r-\Delta r/2-R_x}^{r+\Delta r/2+R_x} \frac{9 x^2 F(x) \mathrm{d}x}{4\pi R_x^3  \left[\left(r+\frac{\Delta r}{2}\right)^3- \left(r-\frac{\Delta r}{2}\right)^3\right]} \left[V\left(R_x, r+\frac{\Delta r}{2}, x \right) - V\left(R_x, r-\frac{\Delta r}{2}, x \right)\right]\right) \nonumber \\
					&= f_i (1 + G_i(r,R_x))
		\end{align}
	Hence, we can get the probability of there being an ionized region at a distance $r$ from a center of an ionized bubble, if we have non-zero  correlation between ionization bubble centers. Using the same method as above,  we can also get the probability of there being a heated region at distance $r$ from a center of an ionized bubble given the correlation between bubble centers. 
		\begin{align}
				P_h = f_i (1 + G_h(r,R_x,R_h))
		\end{align}

\begin{figure}
	\centering
	\begin{minipage}{0.49\textwidth}
		\centering
		\includegraphics[width=0.99\textwidth]{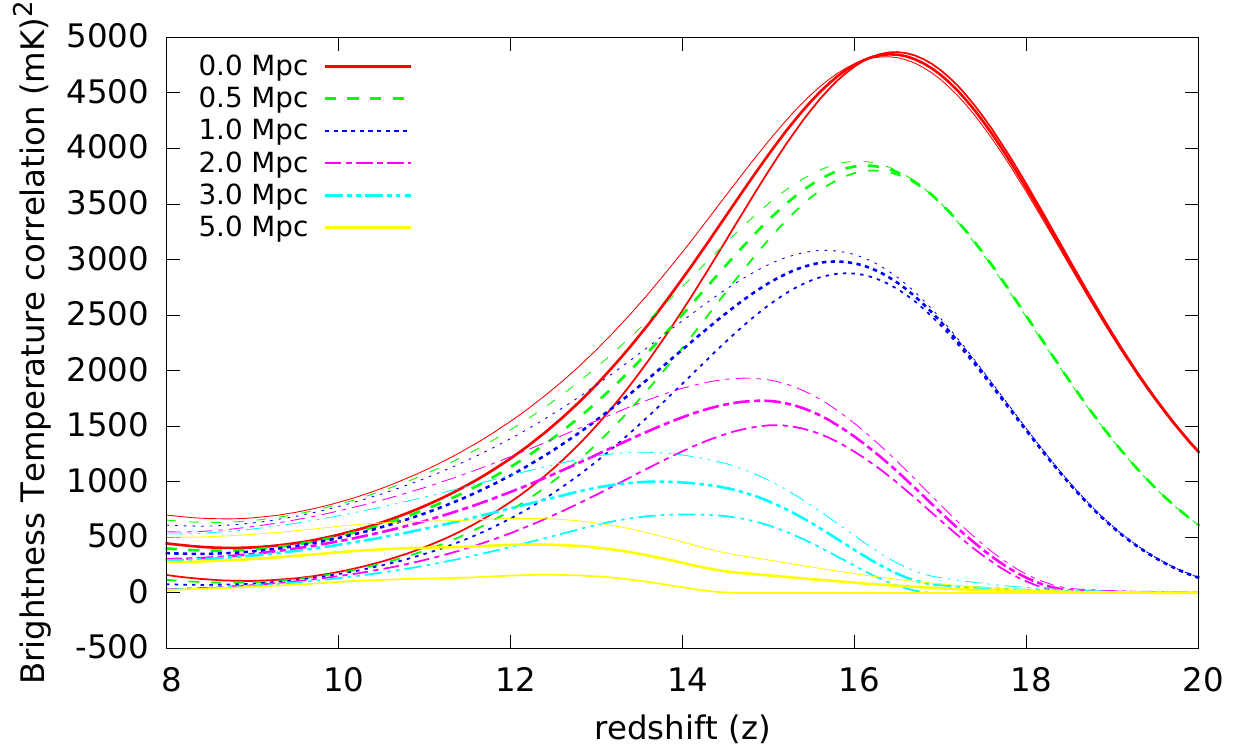}
		\caption{The evolution of correlation function  is displayed for a set of scales for a model in which the background temperature is held constant. Here the curves from bottom to top show cases for no correlation, correlation with $b=1$ and correlation with $b=2$.}
		\label{fig:flat_pro_merge_corr}
	\end{minipage}\hfill
\end{figure}

\subsection{A Simple Model: One bubble size, flat heating profile} \label{sec:corr_simple}
	To analyse the effect of correlation between ionization bubble ceters, we can again study the simple model introduced in section~\ref{sec:flatpro}. Due to correlation of bubble centers, the overlap of bubbles increases, which decreases the heated volume fraction. It also modifies the expression for total correlation:
	\begin{align}
        \mu &= (\psi_h^2 f_h + \psi_b^2 f_b) - (f_h \psi_h + f_b \psi_b)^2 \nonumber \\
				&\quad + \int_{R_h}^{R_h+r} \left((\psi_h-\psi_b)^2 f_h(1- f_i(1+G_i(y)) - f_h (1+ G_h(y))) \frac{{\rm d}C(r,R_x,R_h,y)}{{\rm d}y} \right.\nonumber \\
				&\quad\quad + \left.(\psi_b^2 f_i(1 - f_i(1+G_i(y)) - f_h(1+G_h(y))) + \psi_h^2 f_i f_h(1+G_h(y))) \frac{{\rm d}C(r,0,R_x,y)}{{\rm d}y}\right) {\rm d} y \nonumber \\
				&\quad + \int_{R_x}^{R_h} \psi_h^2  f_i(1-f_i(1+G_i(y))) \frac{{\rm d}C(r,0,R_x,y)}{{\rm d}y} {\rm d}y 
			\label{eq:flatpro_wden_corr}
	\end{align}
	{\bf Here we assume that the correlation of ionization bubble centers follows the same form as the density auto-correlation $F(r)=b\xi(r)$ with a constant bias $b$. We take two possible values of bias: $b=1$ (no bias) and $b=2$}. Figure~{\ref{fig:flat_pro_merge_corr}}	shows the modified correlation for these two cases for the model considered in Figure~{\ref{fig:flat_pro_merge}}. 
	We notice that the  HI signal   increases  due to the correlation of ionization bubble centers. This effect is more significant  at later times 
and on  larger scales. The effect of ionization bubble centers might also introduce correlations on scales at which the signal  would be very small or zero without bubble center correlations, e.g. 
correlation at $r = 5 \, \rm Mpc$ in Figure~{\ref{fig:flat_pro_merge_corr}}.

\end{document}